\documentclass[a4paper,11pt]{article}
\pdfoutput=1 
\usepackage{jheppub} 

\usepackage[T1]{fontenc}

\title{\boldmath Collision of two general particles around a rotating regular Hayward's black holes}

\author[a,1]{Muhammed Amir,\note{Corresponding author.}}
\author[a]{Fazlay Ahmed}
\author[a,b]{and Sushant G. Ghosh}

\affiliation[a]{Centre for Theoretical Physics,
 Jamia Millia Islamia,  New Delhi 110025, India}
\affiliation[b]{Astrophysics and Cosmology Research Unit,
 School of Mathematics, Statistics \\ and Computer Science,
 University of KwaZulu-Natal, Private Bag X54001,\\
 Durban 4000, South Africa}

\emailAdd{amirctp12@gmail.com}
\emailAdd{fazlay@ctp-jamia.res.in}
\emailAdd{sghosh2@jmi.ac.in}

\abstract{The rotating regular Hayward's spacetime, apart from mass ($M$) and angular momentum ($a$), has an additional deviation parameter ($g$) due to the magnetic charge, which generalizes the Kerr black hole when $g\neq0$, and for $g=0$, it goes over to the Kerr black hole. We analyze how the ergoregion is affected by the parameter $g$ to show that the area of ergoregion increases with increasing values of $g$. Further, for each $g$, there exist critical $a_E$, which corresponds to a regular extremal black hole with degenerate horizons $r=r^E_H$, and $a_E$ decrease whereas $r^E_H$ increases with an increase in the parameter  $g$. Ban{\~a}dos, Silk and West (BSW) demonstrated that the extremal Kerr black hole can act as a particle accelerator with arbitrarily high center-of-mass energy ($E_{CM}$) when the collision of two particles takes place near the horizon. We study the BSW process for two particles with different rest masses, $m_1$ and $m_2$, moving in the equatorial plane of extremal Hayward's black hole for different values of $g$, to show that $E_{CM}$ of two colliding particles is arbitrarily high when one of the particles takes a critical value of angular momentum. For a nonextremal case, there always exist a finite upper bound for the $E_{CM}$, which increases with the deviation parameter $g$. Our results, in the limit $g \rightarrow 0$, reduces to that of the Kerr black hole.  }

\begin{document} 
\maketitle
\flushbottom

\section{Introduction}
Recently, Ban{\~a}dos, Silk and West (BSW) \cite{Banados:2009pr} demonstrated that a Kerr black hole can act as a particle accelerator, i.e., when two particles collide arbitrarily close to the  horizon of  an extremal Kerr black hole, the center-of-mass energy  ($E_{CM}$) can grow infinitely in the limiting case of maximal rotation.  The $E_{CM}$  becomes higher for an extremal black hole when one of the colliding particles has a critical angular momentum. This provoked much attention \cite{Banados:2009pr,Jacobson:2009zg,Lake:2010bq,Wei:2010vca,Zaslavskii:2010aw,Zaslavskii:2010jd,
	Zaslavskii:2010pw,Mao:2010di,Wei:2010gq,Li:2010ej,Yang:2012we,Abdujabbarov:2011af,Grib:2010xj,Grib:2011ph,Grib:2010dz,
Patil:2010nt,Patil:2011aw,Patil:2011ya,Patil:2011uf,Harada:2011xz,Liu:2011wv,Liu:2010ja,Harada:2014vka} in the series of subsequent papers by generalizing the BSW mechanism to different spacetimes, also opened a window into new physics with astrophysical applications, e.g., to understand the phenomenon like gamma ray burst and AGNs in Galaxy. Jacabson and Sotiriou \cite{ Jacobson:2009zg} elucidated the BSW mechanism to  discuss some practical limitations for the Kerr black hole to act as a particle accelerator. It was pointed out that infinite $E_{CM}$  for the colliding particles can only be attained when the black hole is exactly extremal and only in infinite time and on the horizon of the black hole. On the other hand, Lake \cite{Lake:2010bq} studied BSW mechanism near the Cauchy horizon to show that $E_{CM}$ of the colliding particles is infinite. When BSW mechanism extended to the Kerr-Newman black hole case \cite{Wei:2010vca}, an unlimited $E_{CM}$ requires an additional restriction on the value of the spin parameter $a$. Zaslavskii \cite{Zaslavskii:2010aw} argued that the similar effect exists for a nonrotating but charged black hole even for the simplest case of radial motion of particles in the Reissner-Nordstr{\"o}m background and also demonstrated that BSW mechanism is a universal property of the rotating black holes \cite{Zaslavskii:2010jd, Zaslavskii:2010pw}. The BSW mechanism has been extended to Einstein-Maxwell-Dilaton black hole \cite{Mao:2010di}, rotating stringy black hole \cite{Wei:2010gq}, Kerr-(anti-) de Sitter Black Hole \cite{Li:2010ej}, BTZ \cite{Yang:2012we}, rotating black hole in Horava-Lifshitz gravity \cite{Abdujabbarov:2011af} and around the four-dimensional Kaluza-Klein extremal black hole in \cite{Mao:2010di}, and it resulted infinitely large $E_{CM}$ near the horizon.

Later, Gao and Zhong showed that the BSW mechanism is possible for the nonextremal black holes to show that for a  critical angular momentum, the $E_{CM}$ diverges at the inner horizon. The interesting work of Grib and Pavlov \cite{Grib:2010xj,Grib:2011ph,Grib:2010dz} shows that the scattering energy of the particles in the centre-of-mass frame can be arbitrarily large not only for extremal black holes but also for nonextremal ones, when we take into account the multiple scattering. It turns out that the divergence of the $E_{CM}$ of the colliding particles is a phenomenon not only associated with black holes but also with naked singularities \cite{Patil:2010nt,Patil:2011aw,Patil:2011ya,Patil:2011uf}. The BSW mechanism was further extended to the case of two different massive colliding particles near the Kerr black hole \cite{Harada:2011xz} and also in the case of Kerr-Newman black hole \cite{Liu:2011wv}, and for the Kerr-Taub-NUT spacetime \cite{Liu:2010ja} suggest that $E_{CM}$ depends not only on the rotation parameter $ a $ but also on the NUT charge $n$. These works on two different massive colliding particles are generalization of previous studies (see also Harada \& Kimura 2014 for a review BSW mechanism \cite{Harada:2014vka}).  

While we are far away from any robust  quantum theory of gravity that tell us how the singularities of classical black holes are solved, there are some models of black hole solutions without the central singularity. These are regular black holes first proposed by Bardeen metric \cite{bardeen}, which is a solution of Einstein’s gravity coupled to a nonlinear electrodynamics field \cite{AyonBeato:1998ub}. Another interesting  model is the Hayward black hole metric \cite{Hayward:2005gi} and Ay\'{o}n-Beato-Garc\'{i}a \cite{AyonBeato:1998ub}. The rotating or Kerr-like solutions of the Bardeen and the Hayward metrics have also been obtained \cite{Bambi:2013ufa}. These rotating  black holes violate even the weak  energy condition is violated, but such a violation can be made very small \cite{Bambi:2013ufa}. Astrophysical black holes are more interested in the study of rotating regular black hole, but the actual nature of these objects has still to be tested \cite{Bambi:2011mj,Bambi:2013qj}. These regular black holes have an additional deviation parameter (say $g$) apart from the mass ($M$) and angular momentum ($a$), which provides a deviation from  the Kerr black hole. It turns out that these black holes, for each nonzero $g$, there exist critical $a_{E}$ , which corresponds to a regular extremal black hole with degenerate horizons \cite{Ghosh:2014pba}. The study of BSW mechanism for these rotating regular black holes has also been analyzed \cite{Ghosh:2014mea,Amir:2015pja,Ghosh:2015pra} suggesting that a rotating regular black hole can also act as a particle accelerator, which  in turn provides a suitable framework for Plank-scale physics.

In the current paper, we want to discuss the collision of two different massive particles falling from rest at infinity in the background of rotating regular Hayward's black holes. 

\section{Rotating Regular Hayward's Black Holes}
\label{bh}
The spherically symmetric Hayward's metric \cite{Hayward:2005gi} is given by
\begin{equation}\label{mtrc}
ds^2 = -f(r)dt^2+\frac{1}{f(r)}dr^2 +r^2 d\Omega^2,
\end{equation}
with $$f(r) = 1-\frac{2mr^2}{r^3+2l^2m} \;\;\; \text{and} \;\;\;\; d\Omega^2 = d\theta^2 +\sin^2 \theta d\phi^2,$$ where $m$ is mass and $l$ is constant. The metric (\ref{mtrc}) is asymptotically behaves as 
$$f(r) \sim 1-\frac{2m}{r} \;\;\; \text{as} \;\;\; r \rightarrow \infty, $$ 
where near center 
$$f(r) \sim 1-\frac{r^2}{l^2} \;\;\; \text{as} \;\;\; r \rightarrow 0.$$ 
The solution (\ref{mtrc}), for $l=0$, reduces to well known Schwarzschild black hole and is flat for $m=0$. The metric function and the curvature invariants are well behaved everywhere, including at origin. The analysis $f(r)=0$, imply a critical mass $m^* = 3\sqrt{3}l/4$ and critical radius $r^* = \sqrt{3}l$, such that a regular extremal black hole with degenerate horizons $r=r^*$ when $m=m^*$. When $m<m^*$, a regular nonextremal black hole horizon with both Cauchy and event horizon, corresponding to two roots of $f(r)=0$, and no black hole when $m>m^*$.

Next, the rotating spacetime corresponding to (\ref{mtrc}) is also obtained \cite{Bambi:2013ufa}, which looks similar to the Kerr black hole except that $M$ is replaced by $m(r)$. In this section, we shall discuss the horizon structure and the ergoregion of rotating regular Hayward's black hole metric. The rotating regular Hayward's black hole in the Boyer\(-\)Lindquist coordinates reads \cite{Bambi:2013ufa}: 
\begin{eqnarray}
\label{metric}
ds^2 &=& -\left[1-\frac{2m(r)r}{\Sigma}\right]dt^2-\frac{4am(r)r \sin^{2} \theta}{\Sigma}dt d\phi+\frac{\Sigma}{\Delta}dr^2 \nonumber\\ 
&+&\Sigma d \theta^2 +\left[r^2+a^2+\frac{2 a^2 m(r)r \sin^2 \theta}{\Sigma}\right]\sin^2 \theta d \phi^2,
\end{eqnarray}
where $\Sigma$ and $\Delta$, respectively, are given by
\begin{eqnarray}
\label{delta}
\Sigma=r^2 + a^{2}\cos^{2} \theta , \nonumber\\  \Delta =r^2 + a^2 - 2 m(r) r.
\end{eqnarray} 
Here $a$ is a rotation parameter, and $m(r)$ is function related to mass of the black hole via
\begin{eqnarray}
\label{m}
m(r)=M \frac{r^{3+\alpha} \Sigma^{-\alpha/2}}{r^{3+\alpha} \Sigma^{-\alpha/2}+g^3 r^{\beta} \Sigma^{-\beta/2}},
\end{eqnarray}
where $M$ represents black hole mass, $g$ is magnetic charge which provides deviation from the standard Kerr black hole, and $\alpha$, $\beta$ are two real numbers. In the equatorial plane ($\theta = \pi/2$), the mass function (\ref{m}) takes the form
\begin{eqnarray}\label{mas}
m(r)=M \frac{r^3}{r^3+g^3},
\end{eqnarray}
which is independent of parameter $\alpha$, $\beta$ and $\theta$. The metric (\ref{metric}) represents a rotating regular Hayward's black hole \cite{Bambi:2013ufa}, which is a generalization of the Kerr spacetime because when $g=0$, then it reduces to the standard Kerr black hole \cite{Kerr:1963ud} and for both $a=g=0$, the metric reduces into the Schwarzschild black hole \cite{schw}. A Hayward's regular black hole satisfies the weak energy condition only for a nonrotating case ($a=0$) and violates the weak energy condition for rotating case \cite{Bambi:2013ufa}. The metric is regular everywhere, including at $r=0$, $\theta=\pi/2$, as the Ricci scalar ($R_{ab}R^{ab}$) and Kretschmann scalar ($R_{abcd}R^{abcd}$) are well behaved \cite{Amir:2015pja}.
\begin{figure}[tbp]
\centering 
\includegraphics[width=.24\textwidth]{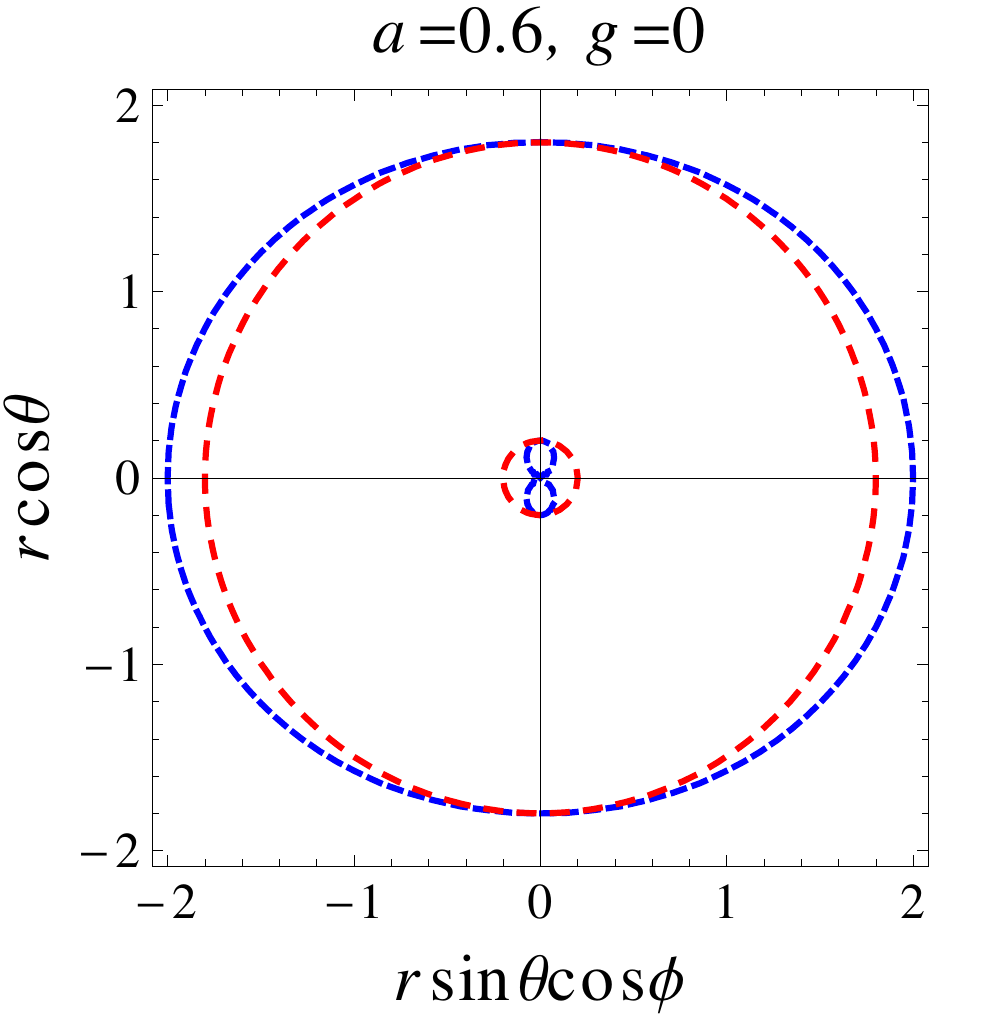}
\hfill
\includegraphics[width=.24\textwidth]{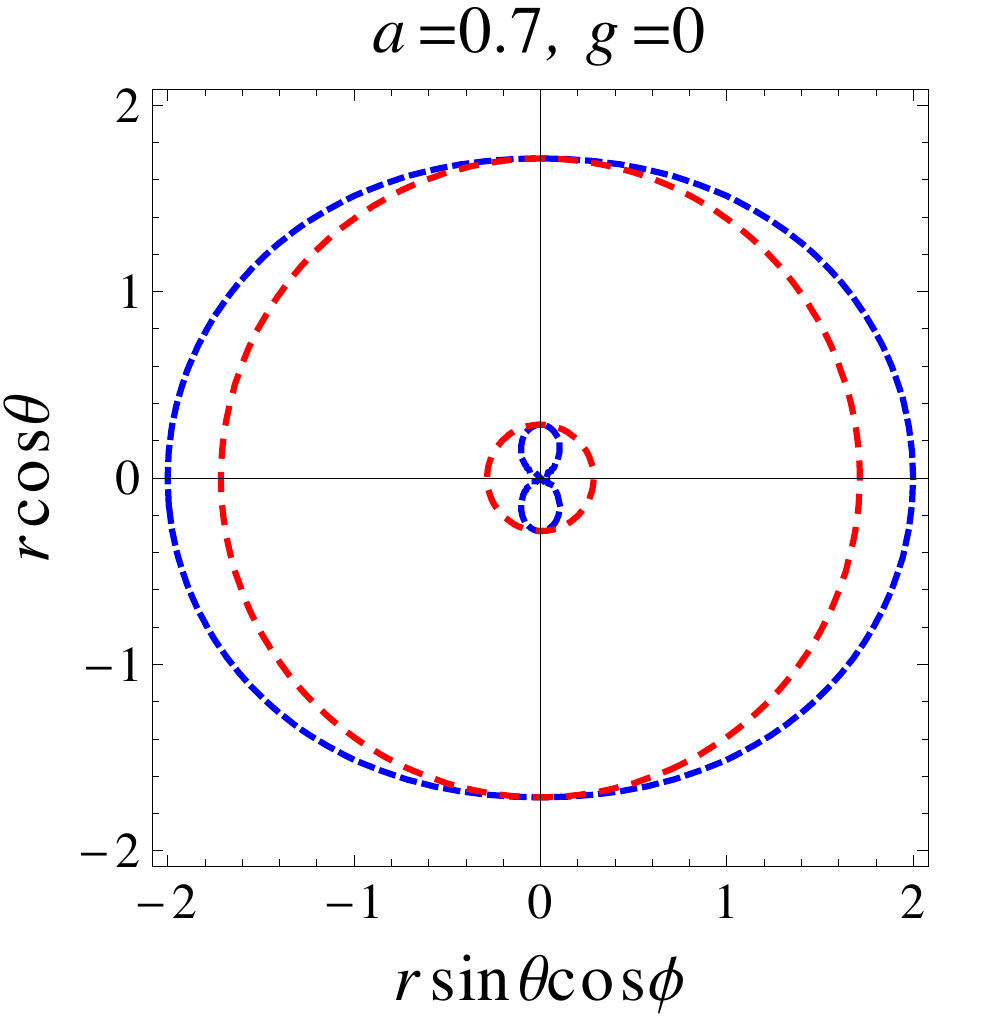}
\hfill
\includegraphics[width=.24\textwidth]{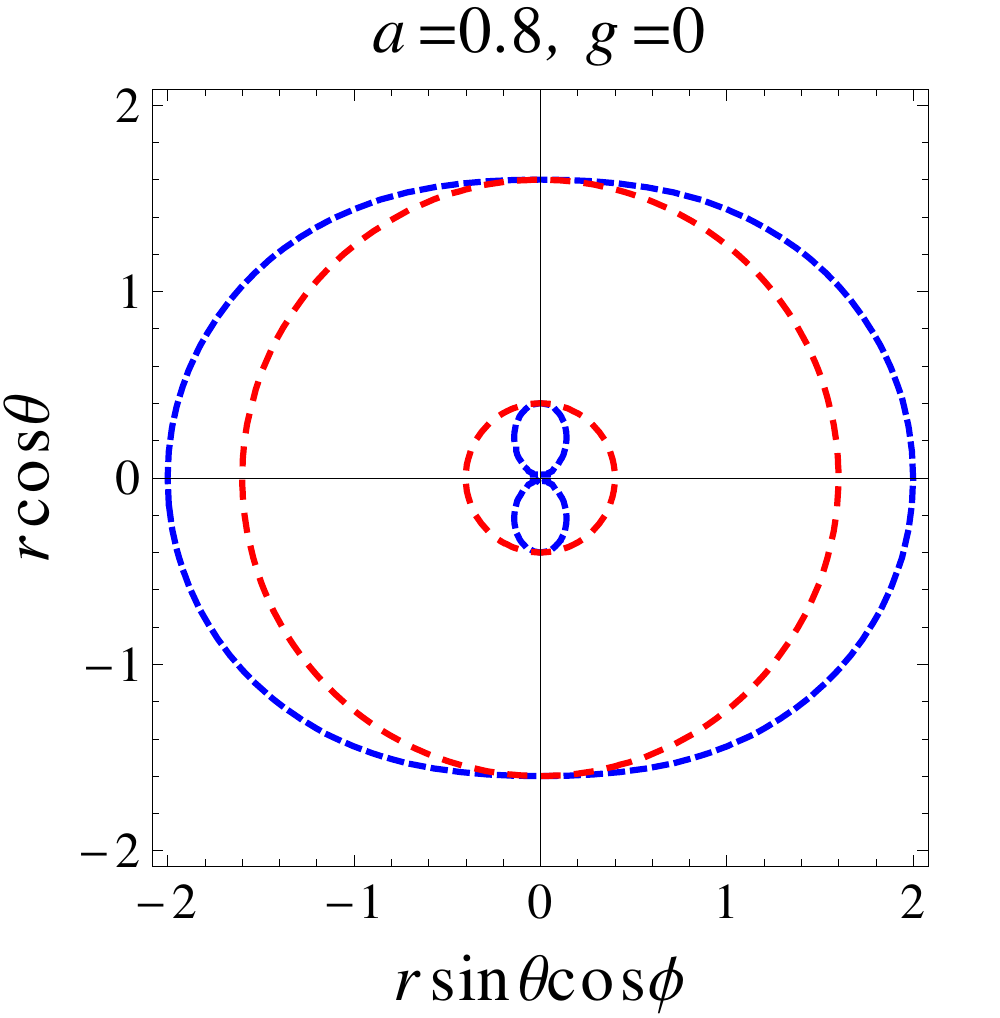}
\hfill
\includegraphics[width=.24\textwidth]{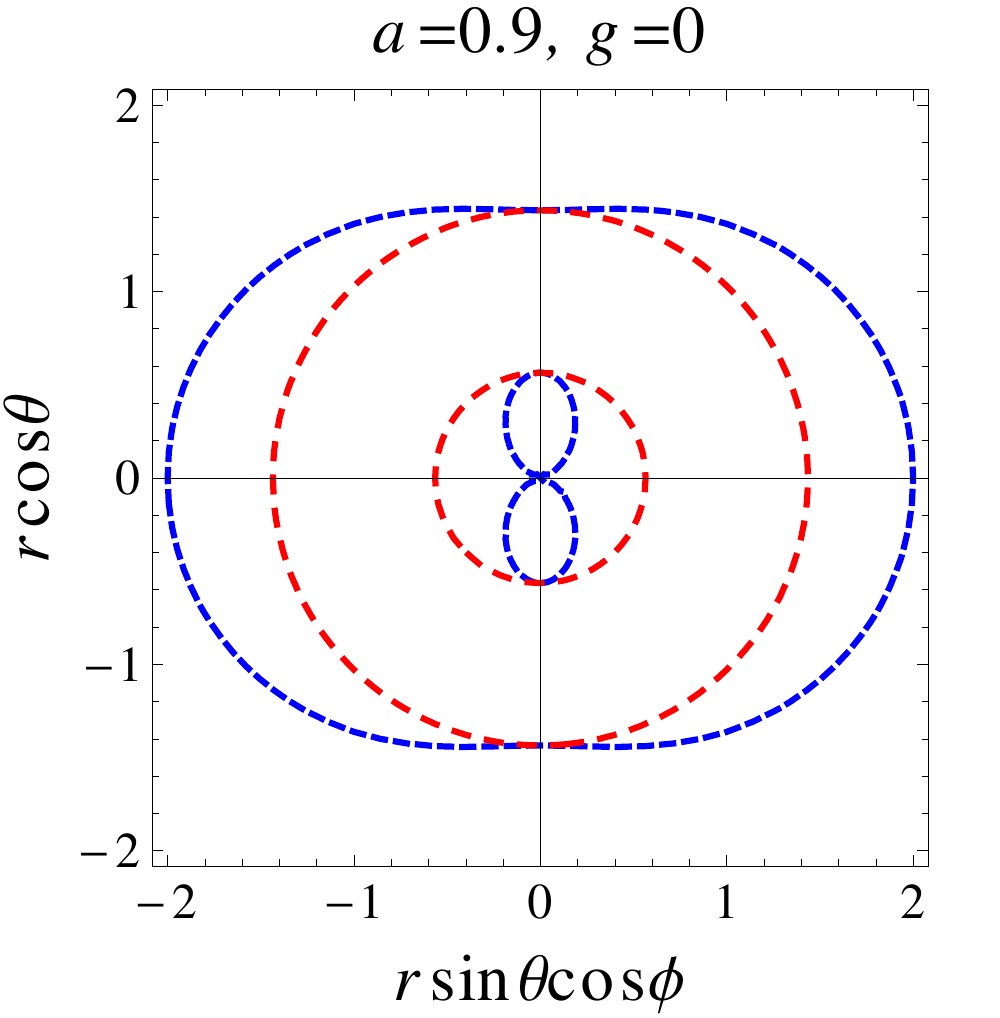}
\caption{\label{fig1} Plot showing the behaviour of ergoregion in the xz-plane of rotating regular Hayward's black hole for $g=0$ with different values of $a$. The blue and red lines correspond to the static limit surface and horizons, respectively.}
\end{figure}
\begin{figure}[tbp]
\centering 
\includegraphics[width=.24\textwidth]{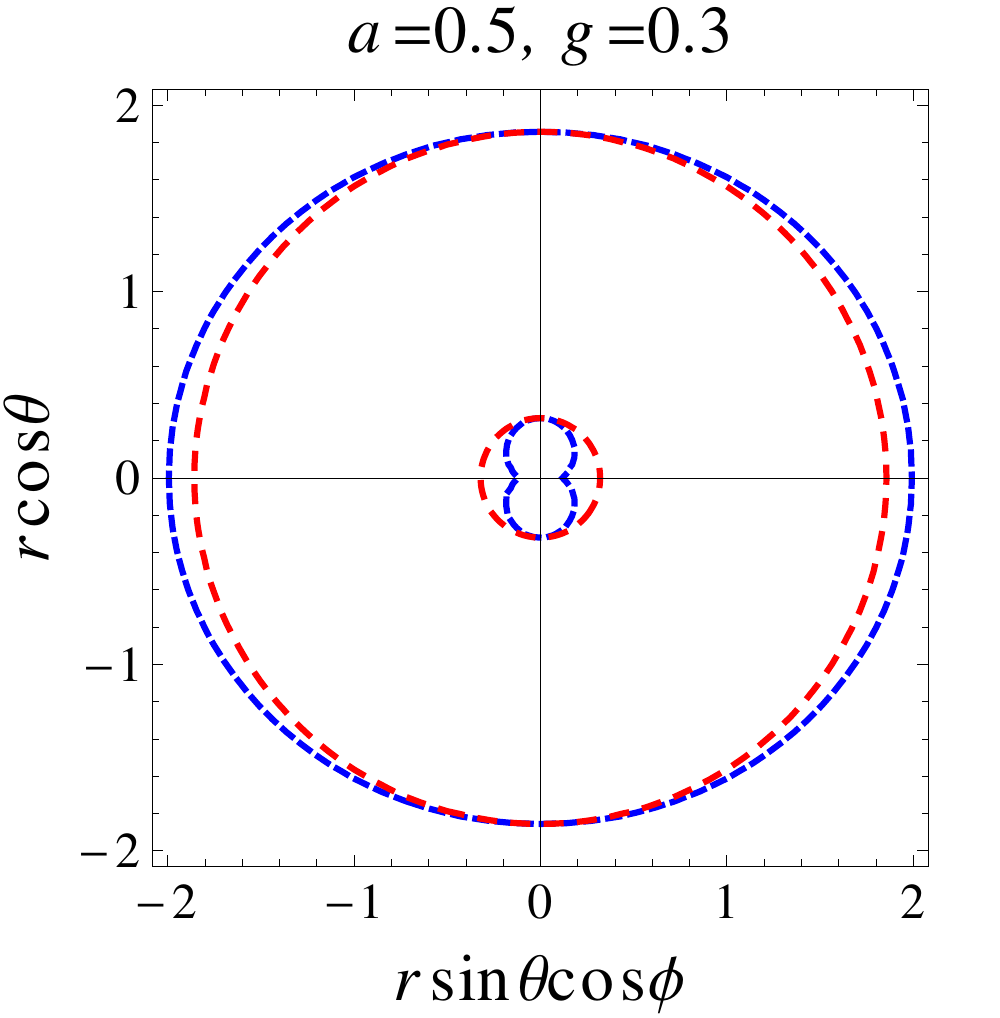}
\hfill
\includegraphics[width=.24\textwidth]{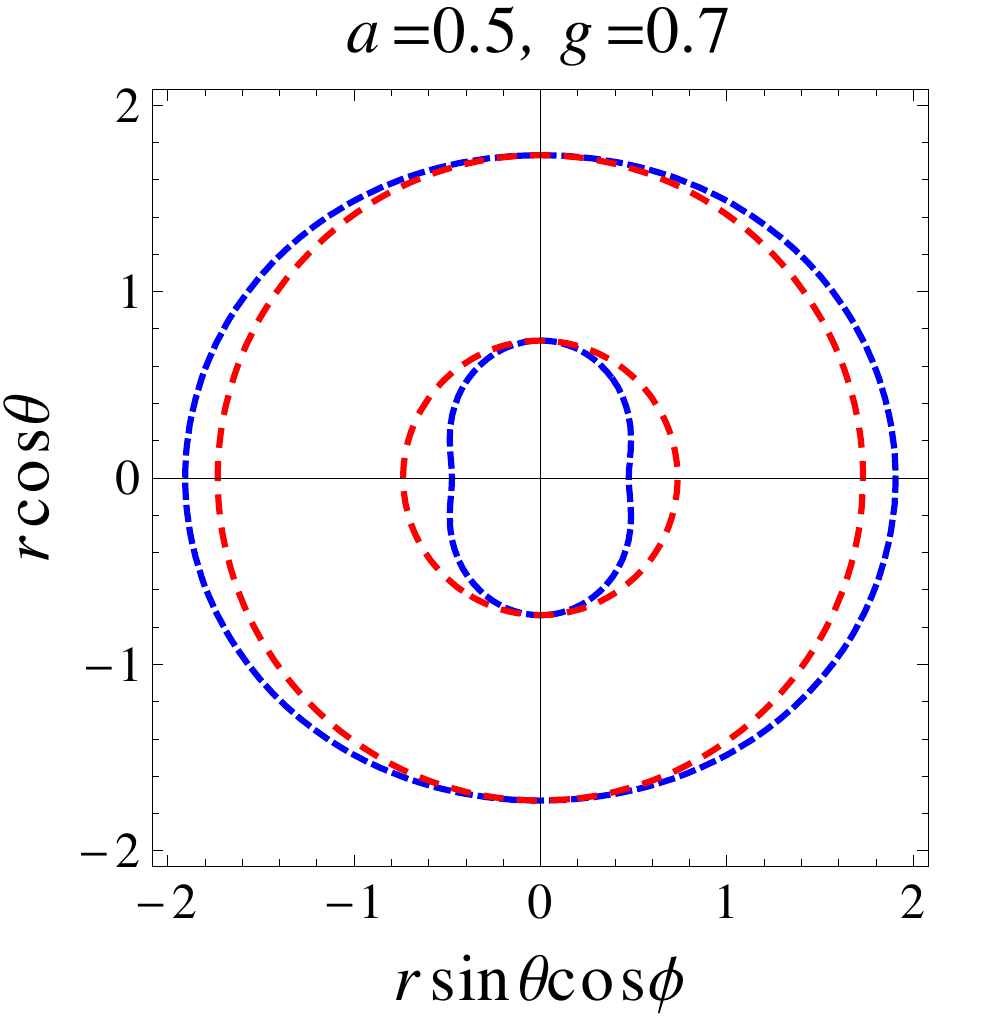}
\hfill
\includegraphics[width=.24\textwidth]{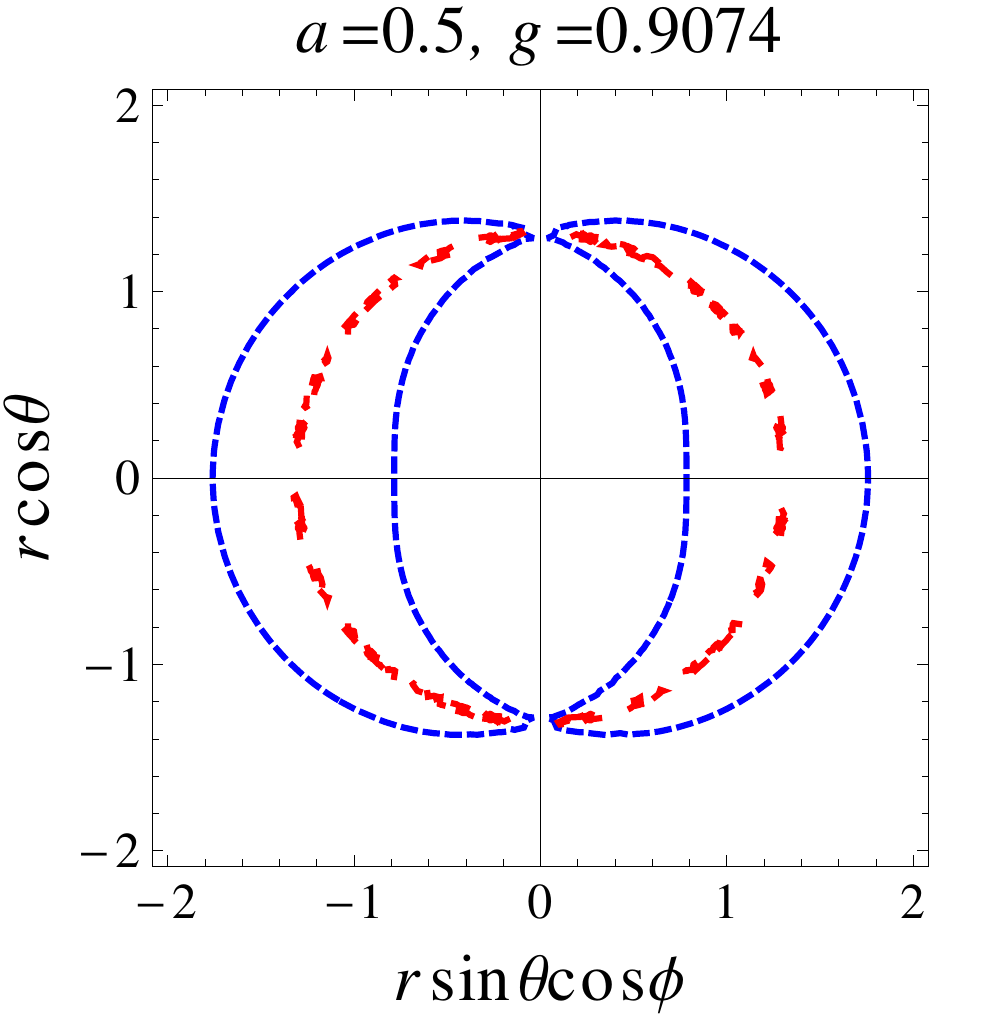}
\hfill
\includegraphics[width=.24\textwidth]{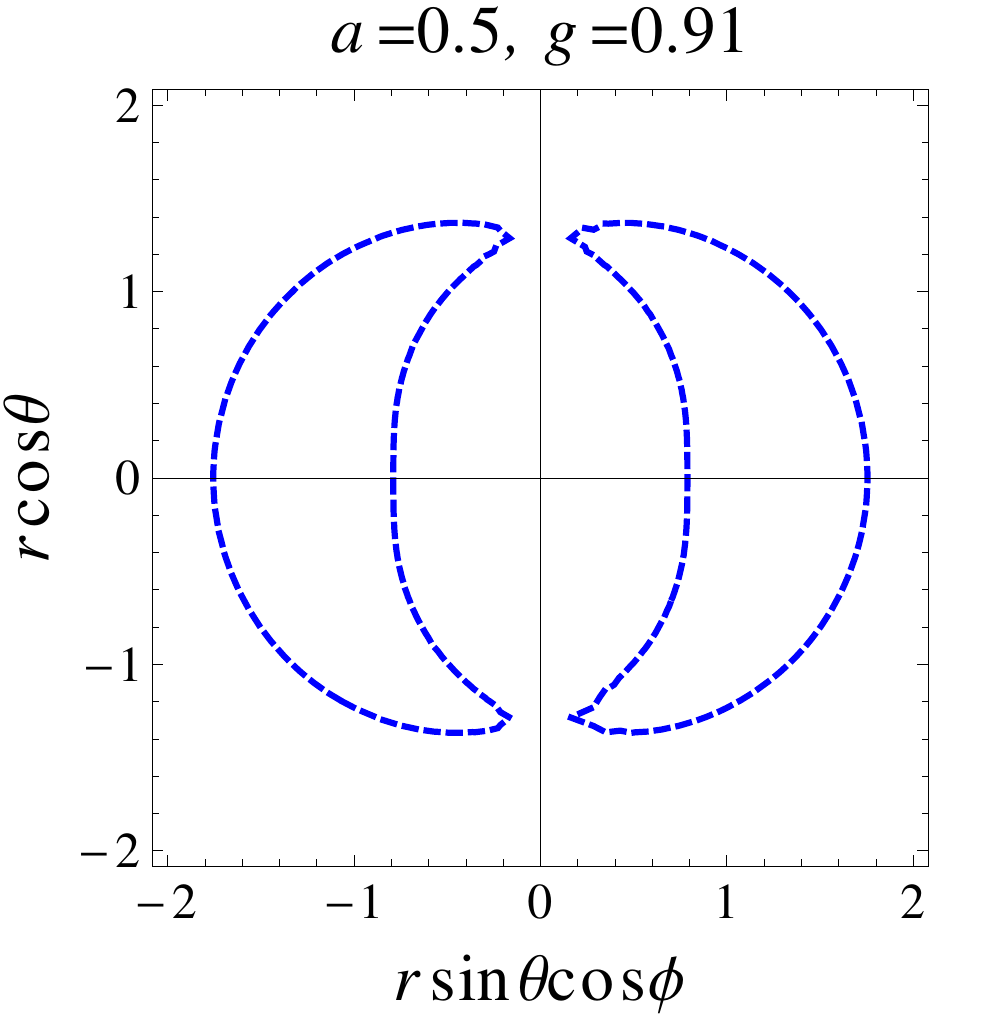}
\\
\includegraphics[width=.24\textwidth]{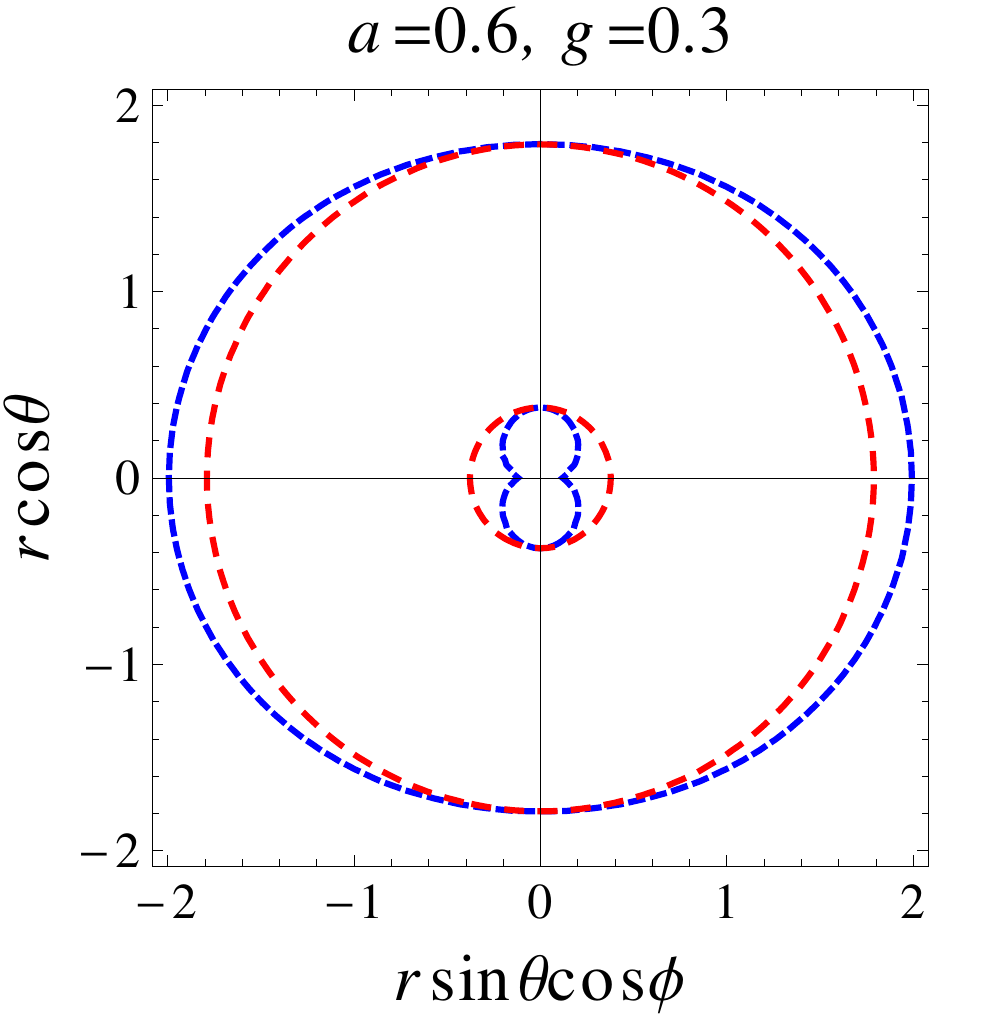}
\hfill
\includegraphics[width=.24\textwidth]{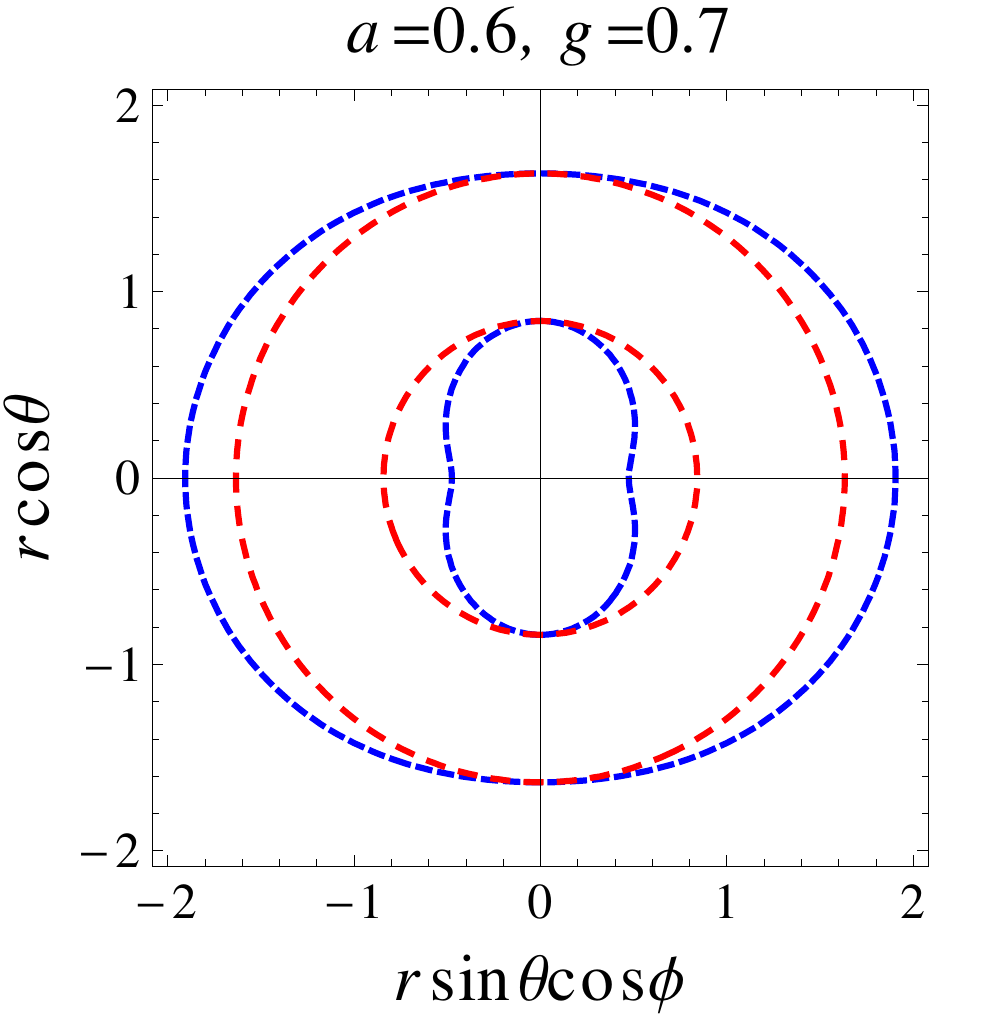}
\hfill
\includegraphics[width=.24\textwidth]{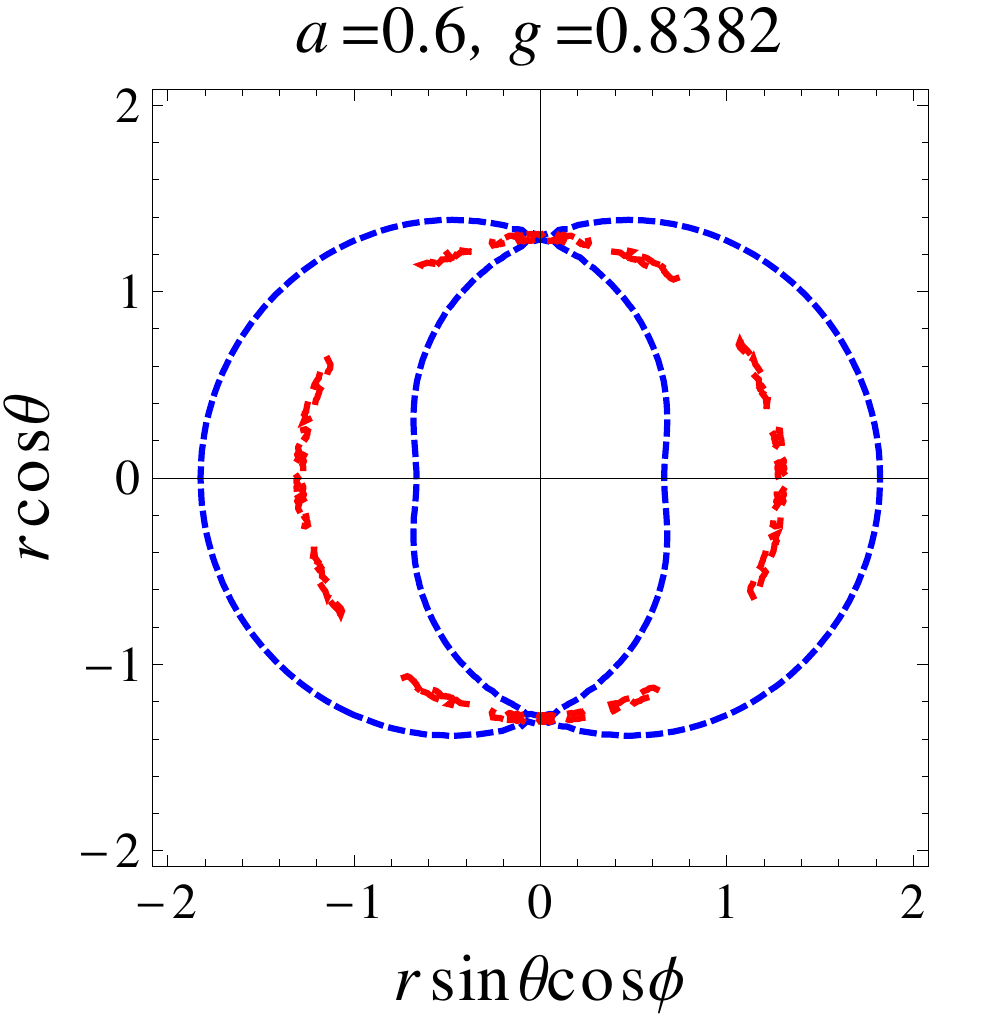}
\hfill
\includegraphics[width=.24\textwidth]{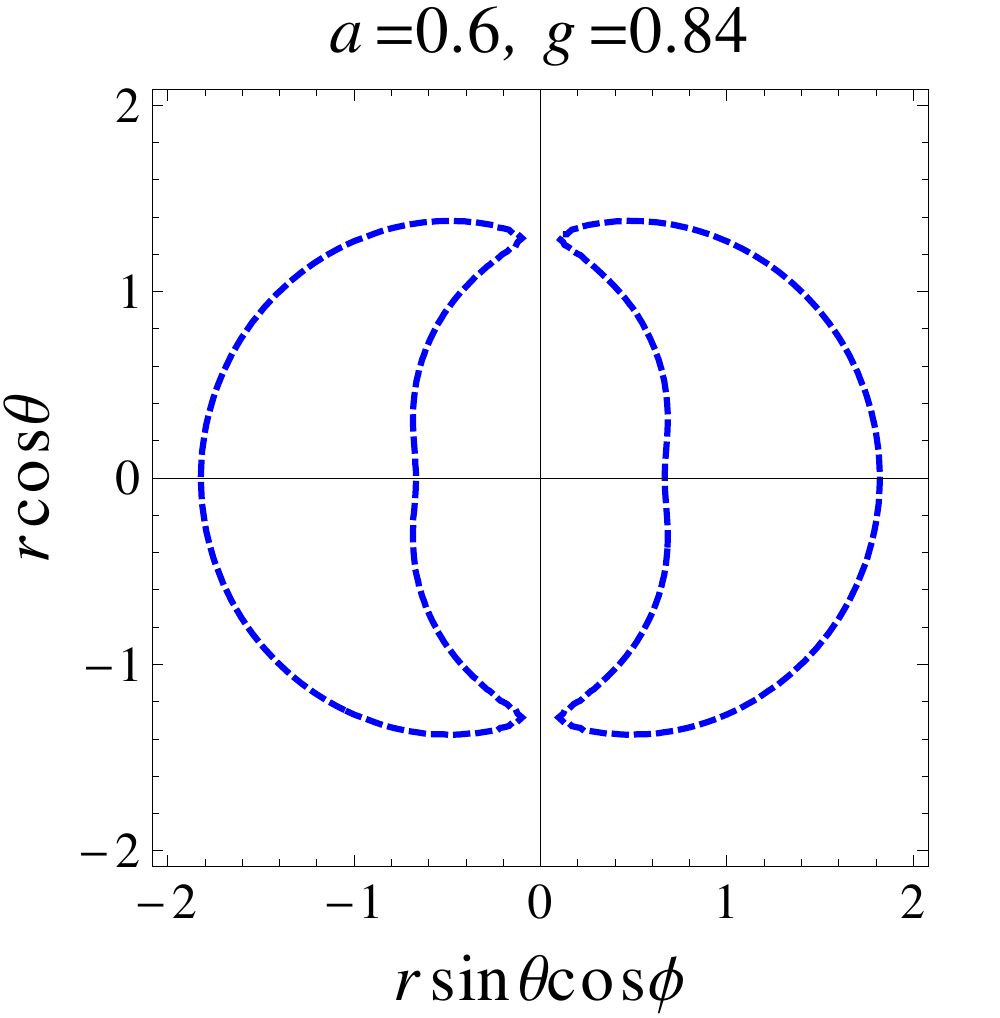}
\\
\includegraphics[width=.24\textwidth]{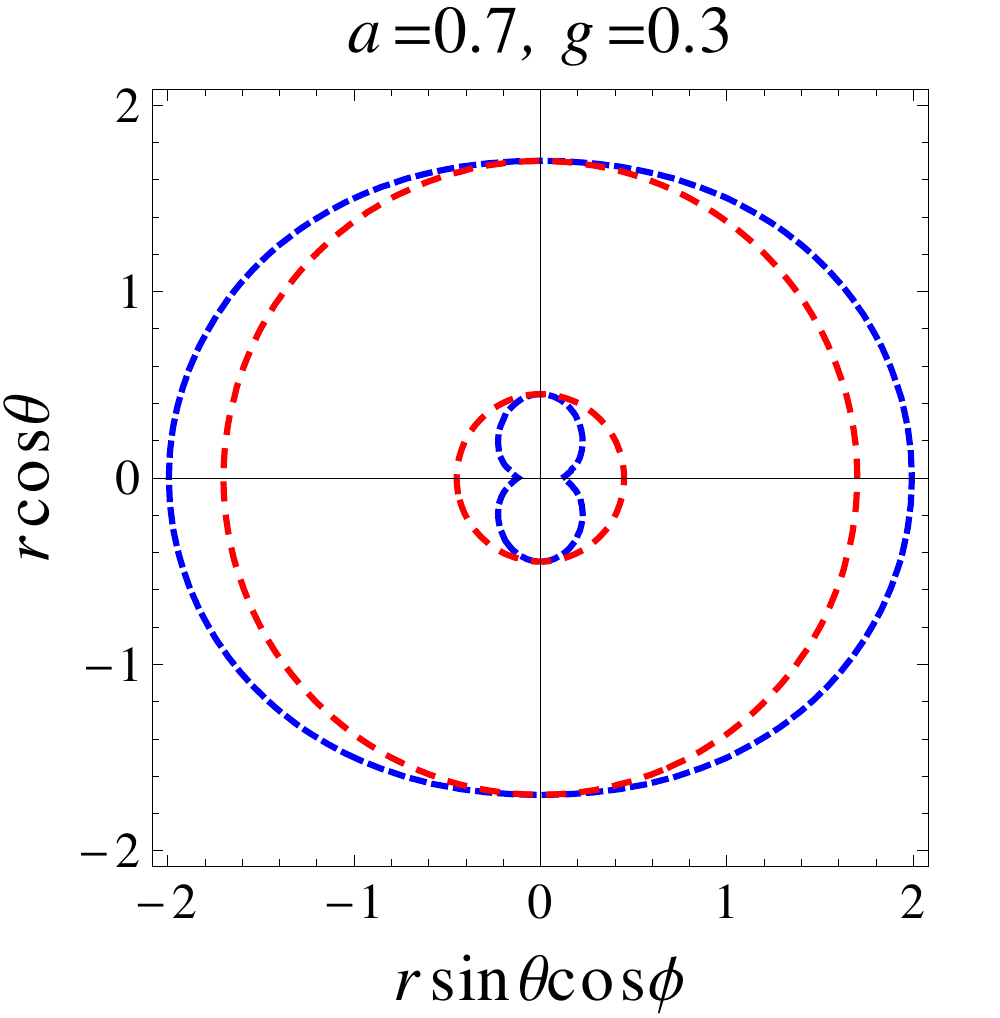}
\hfill
\includegraphics[width=.24\textwidth]{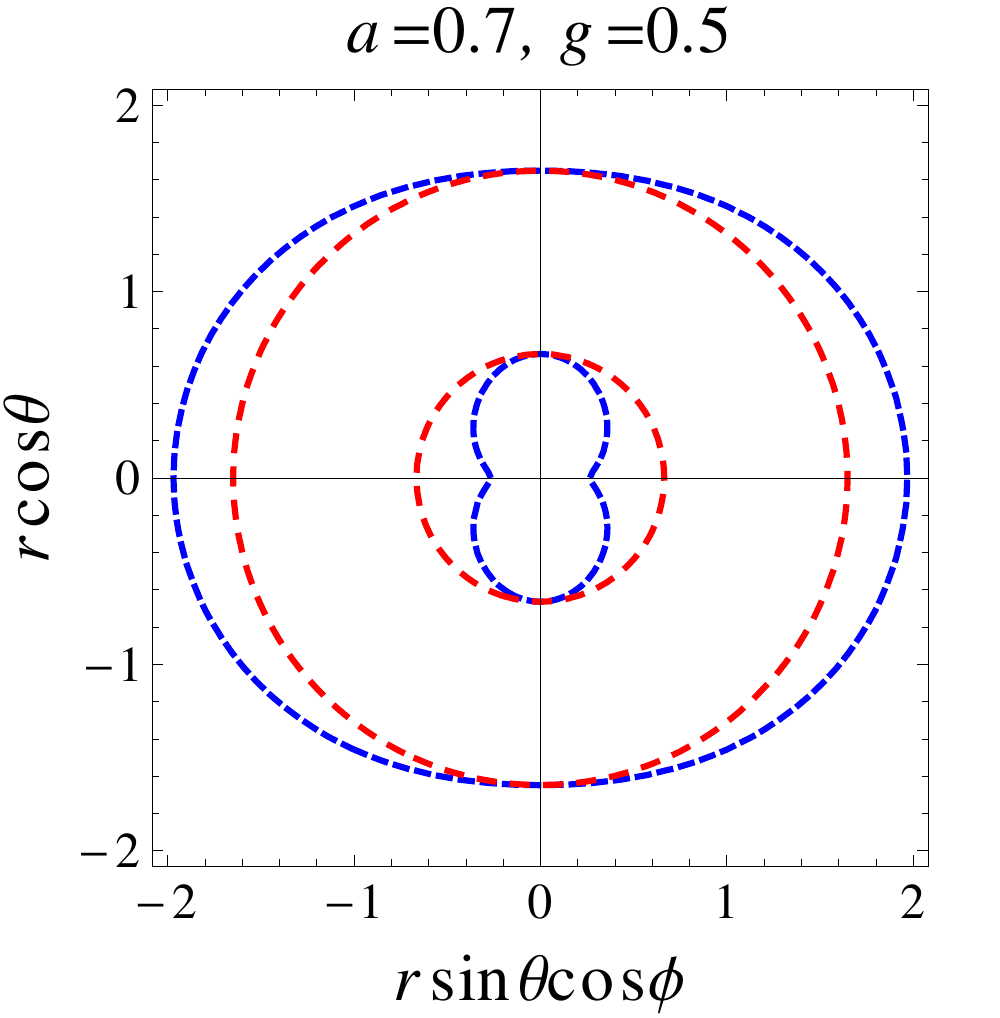}
\hfill
\includegraphics[width=.24\textwidth]{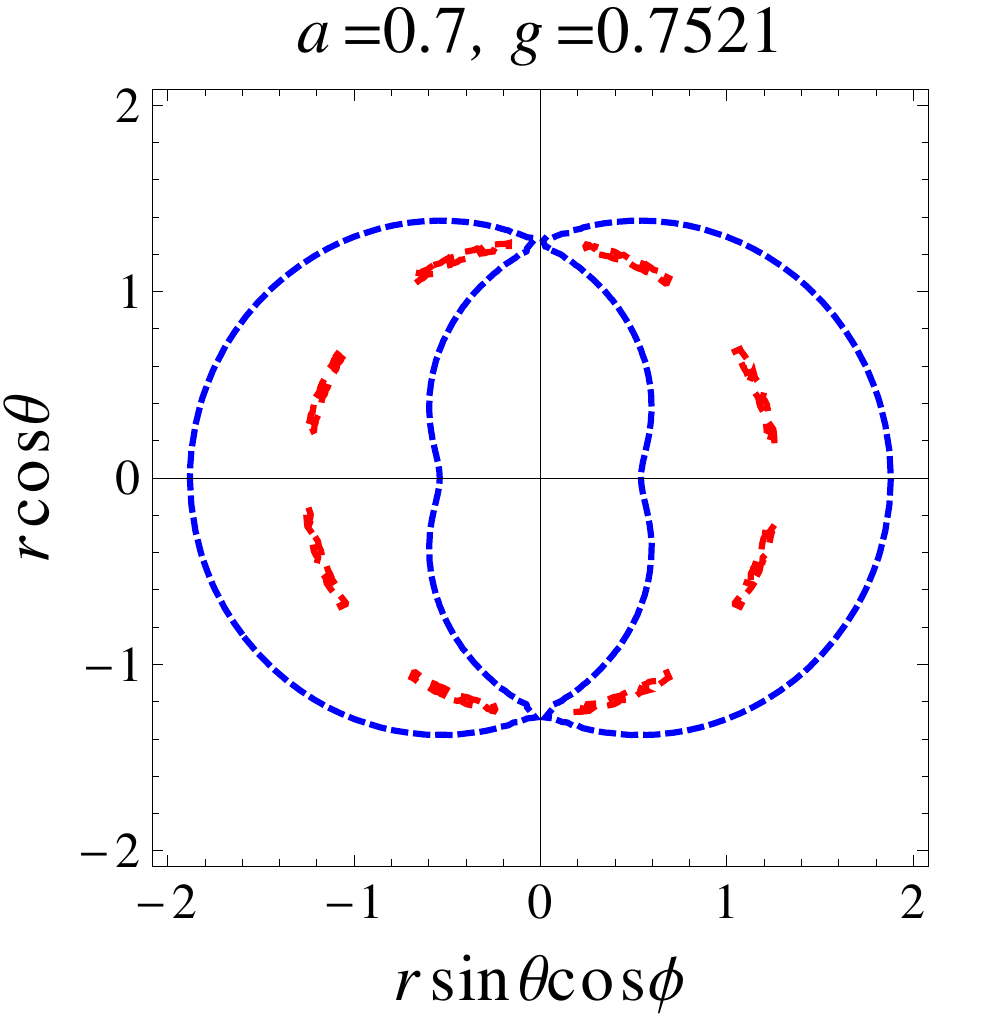}
\hfill
\includegraphics[width=.24\textwidth]{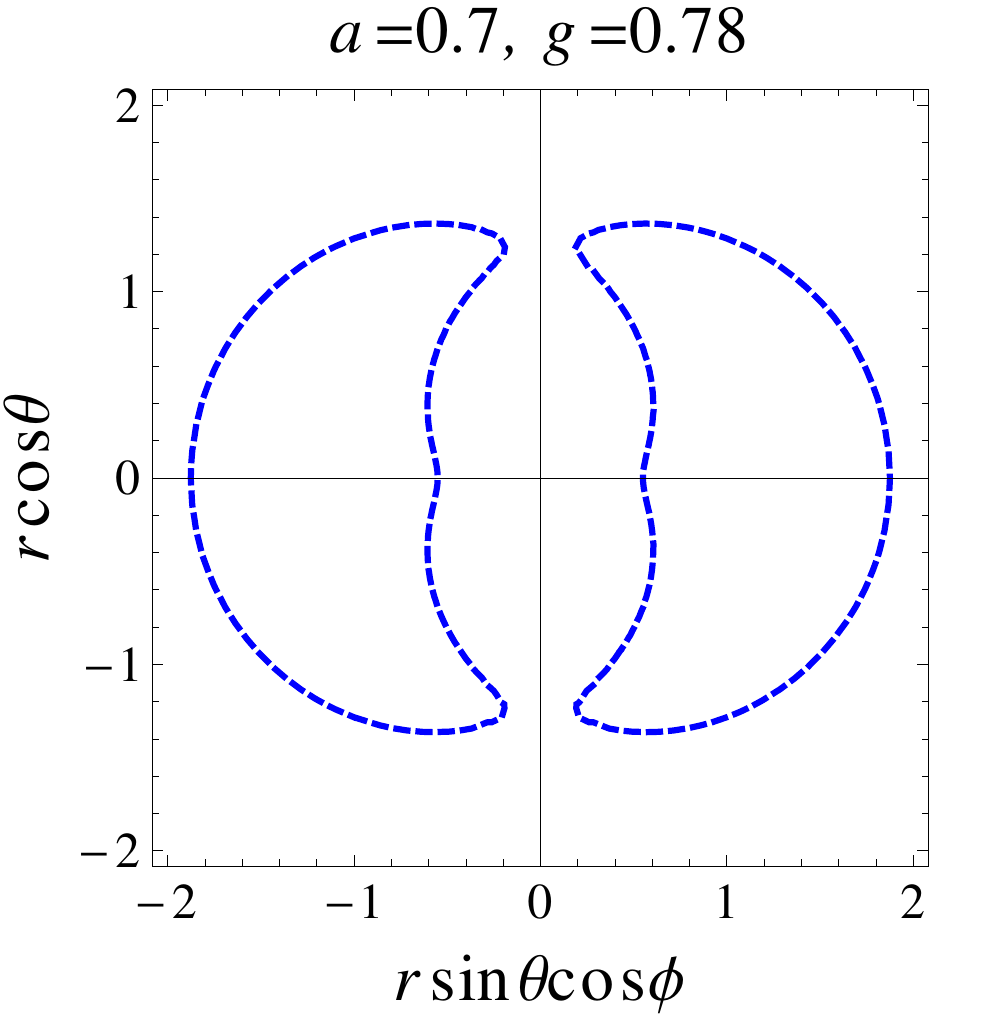}
\\
\includegraphics[width=.24\textwidth]{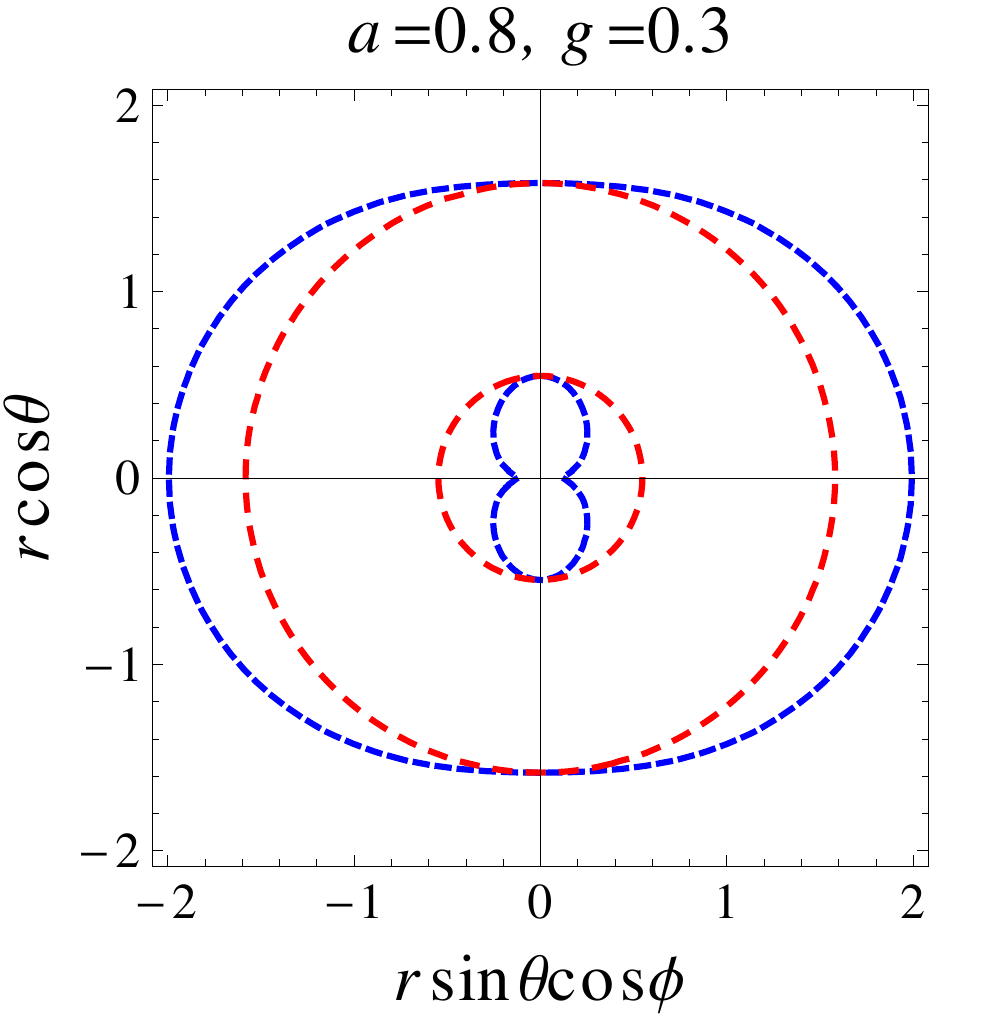}
\hfill
\includegraphics[width=.24\textwidth]{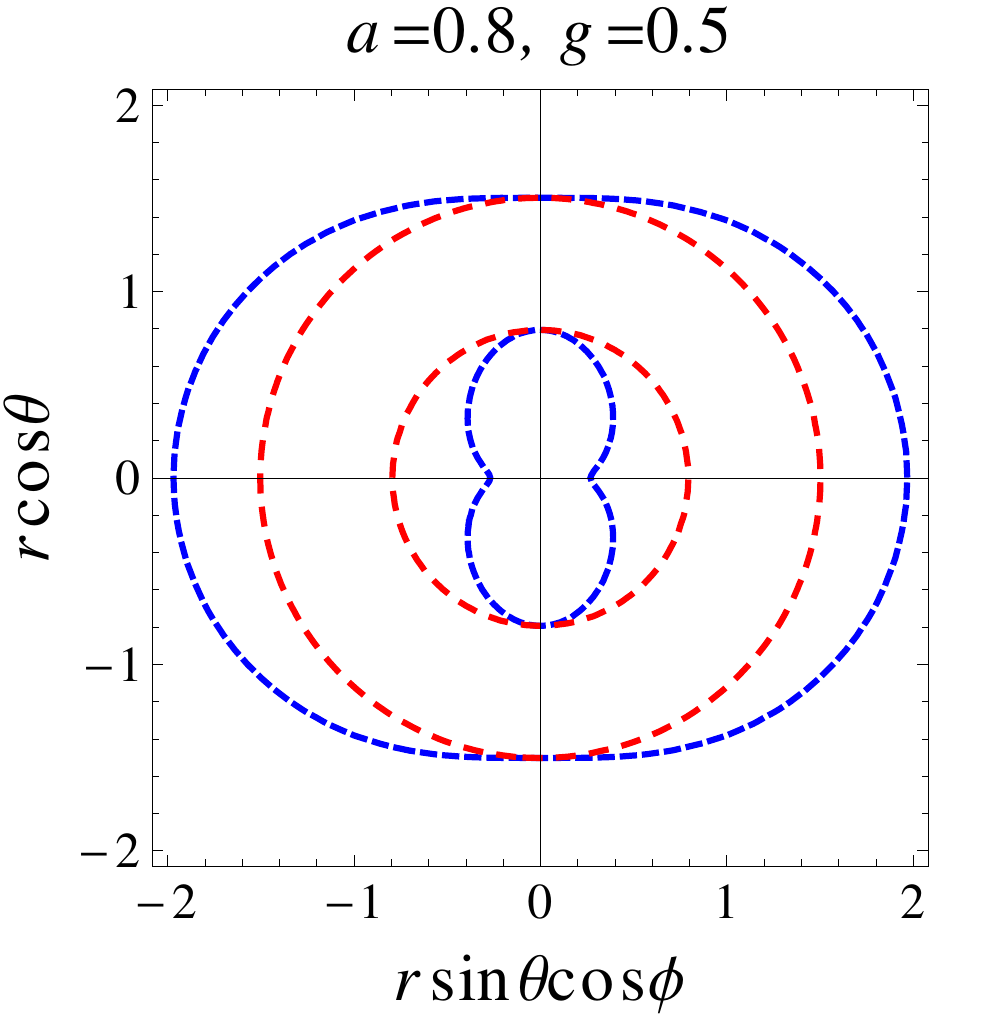}
\hfill
\includegraphics[width=.24\textwidth]{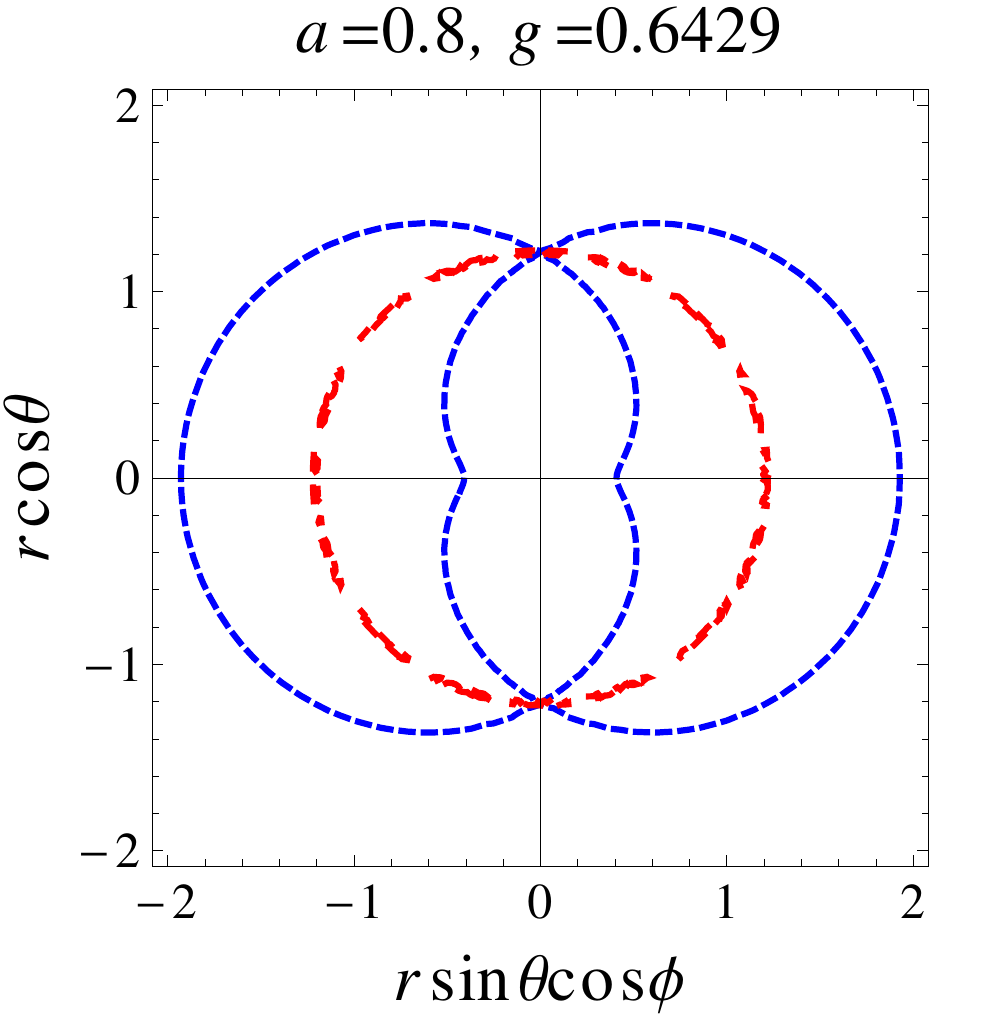}
\hfill
\includegraphics[width=.24\textwidth]{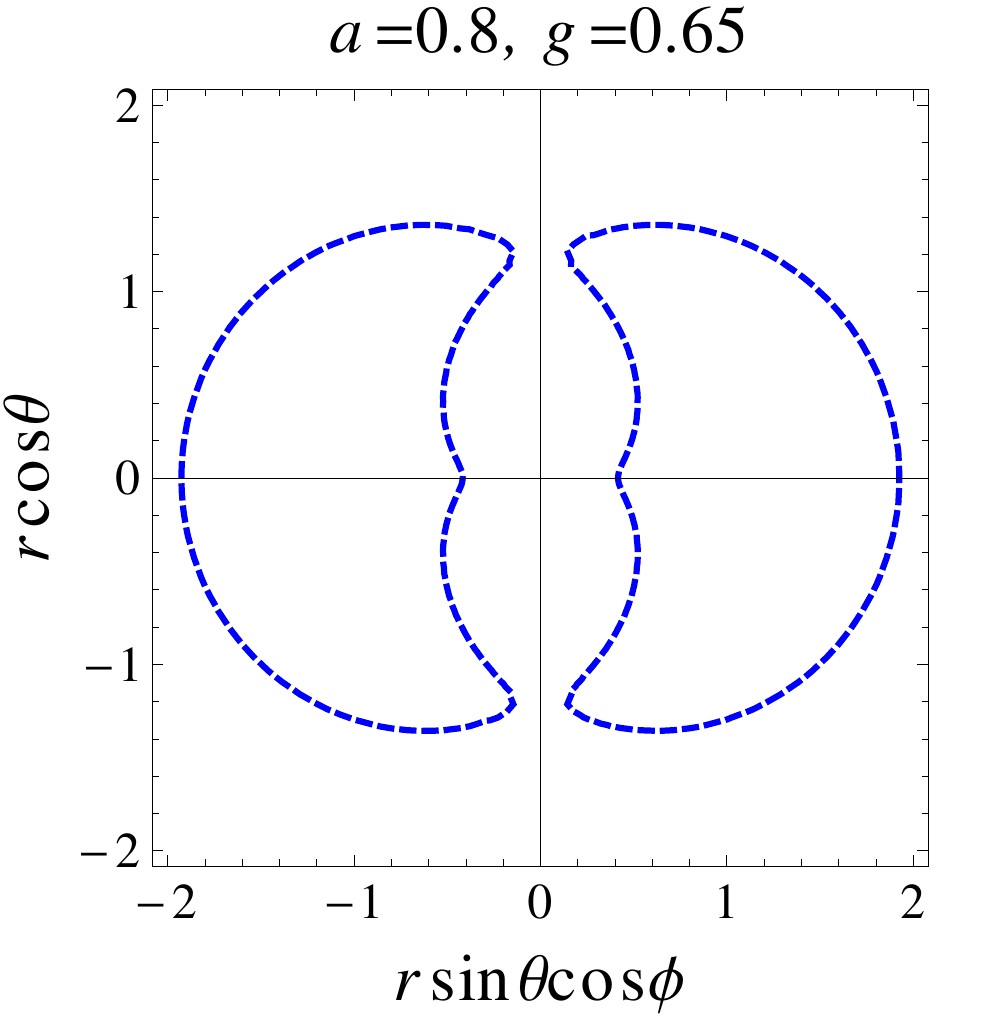}
\\
\includegraphics[width=.24\textwidth]{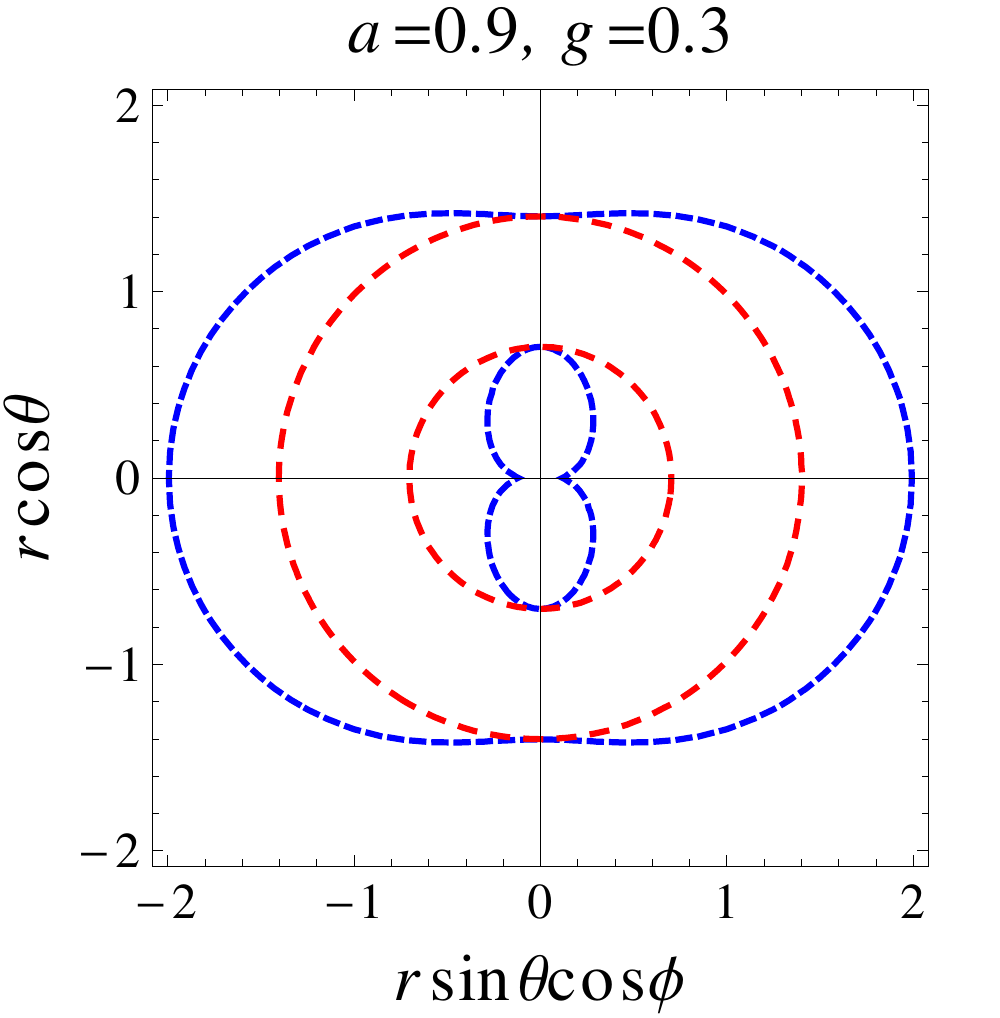}
\hfill
\includegraphics[width=.24\textwidth]{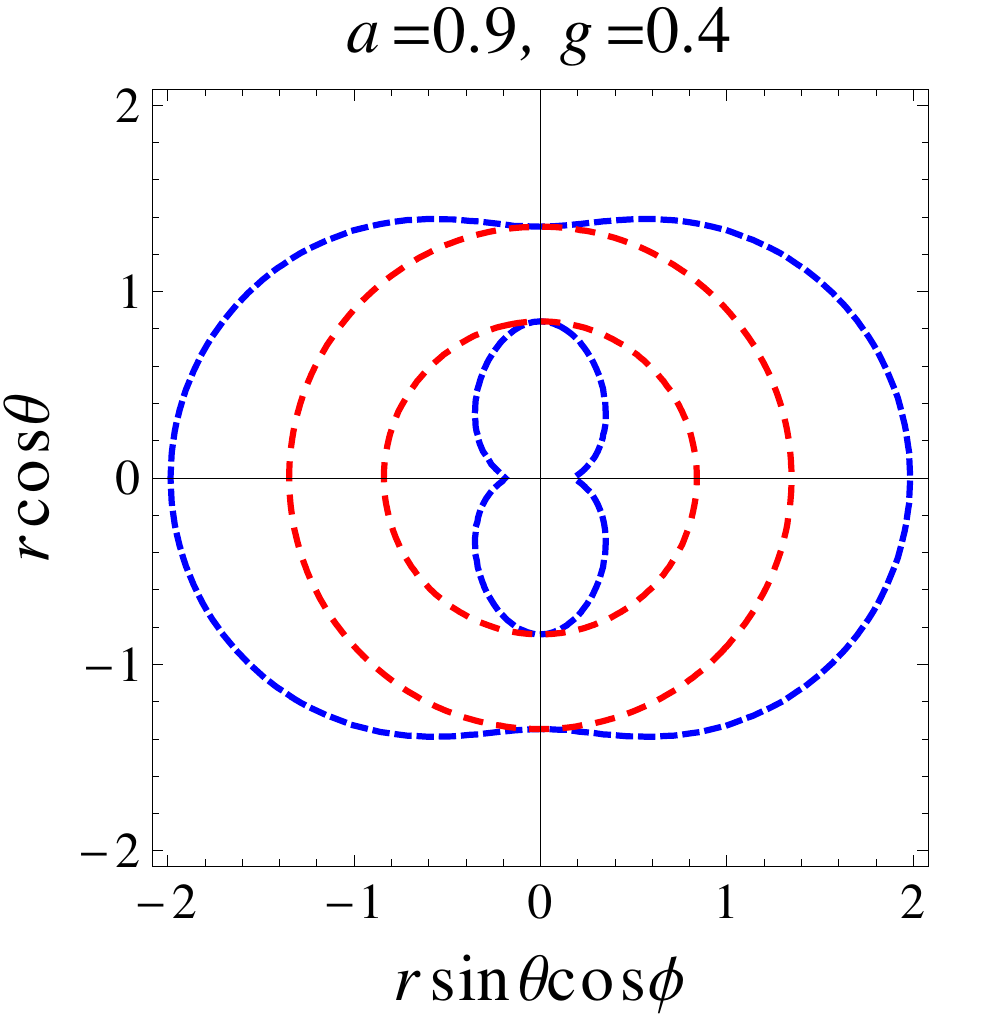}
\hfill
\includegraphics[width=.24\textwidth]{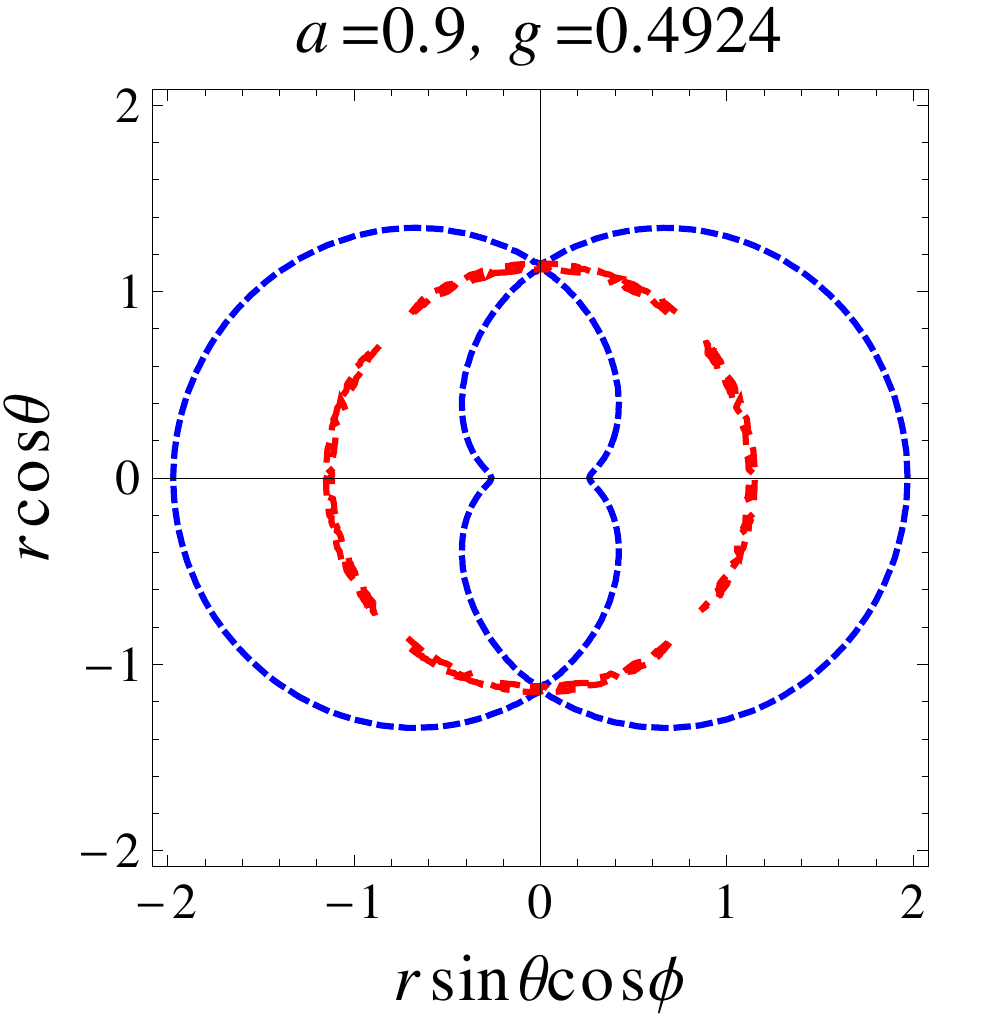}
\hfill
\includegraphics[width=.24\textwidth]{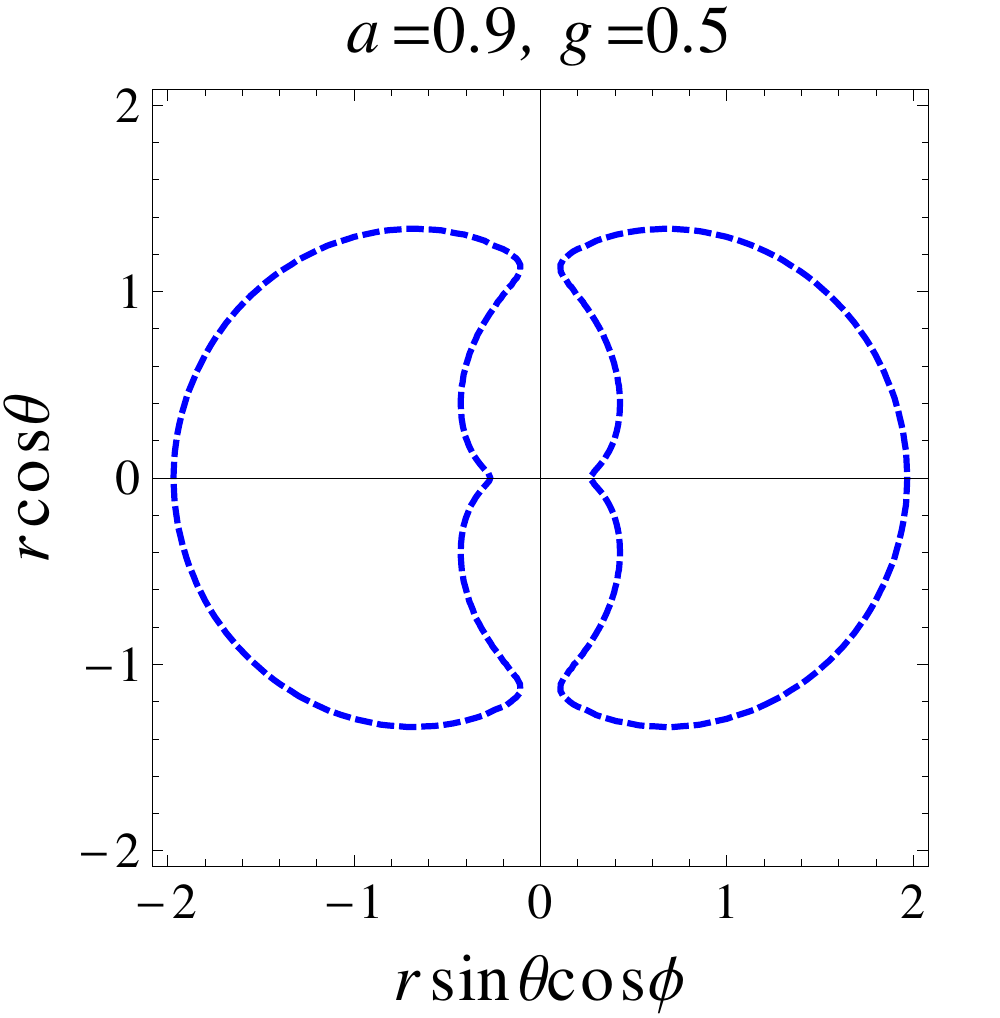}
\caption{\label{fig2} Plot showing the behaviour of ergoregion in the $xz$-plane of rotating regular Hayward's black hole for different values of $g$. Here we take $M=1$, $\theta=\pi/2$. The blue and red lines correspond to the static limit surface and horizons, respectively.}
\end{figure}
\subsection{Ergoregion of rotating regular Hayward's black hole}
Recently, the horizons structure of the rotating regular Hayward's black hole has been analyzed \cite{Amir:2015pja}. Here, we are going to discuss the ergoregion of the rotating regular Hayward's black hole. The metric (\ref{metric}) becomes singular at $\Delta=0$, which is a coordinate singularity. The event horizon of the rotating regular Hayward's black hole is given by zeros of $\Delta=0$,
\begin{eqnarray}\label{hor}
r^2 + a^2 -2 m(r) r=0.
\end{eqnarray}
It is shown that eq.~(\ref{hor}) admits two roots which correspond to, respectively, the event horizon and the Cauchy horizon. Definitely, the event horizon depends on $\theta$, and hence different from the Kerr black hole. If both of the horizons coincide, then we get an extremal black hole otherwise it is known as a nonextremal black hole. Thus, the rotating regular Hayward's black hole has an extremal black hole for each nonzero value of the deviation parameter $g$. There is another important surface of a black hole, particularly known as a static limit surface.  The main property of the static limit surface is that the nature of particle geodesics change after crossing the static limit surface, i.e., timelike geodesics becomes spacelike and vice versa. The static limit surface satisfies $g_{tt}=0$, i.e.,
\begin{eqnarray}
\label{sls}
r^2+ a^2 \cos^{2} \theta- 2 m(r) r =0.
\end{eqnarray}
From eqs.~(\ref{hor}) and (\ref{sls}), it is clear that the horizons and the static limit surface depend on constants $\alpha$, $\beta$, $g$ and $\theta$. It turns out that for each $g$, there exist critical $a_E$ and $r_H^E$, which corresponds to a regular extremal black hole with degenerate horizons, and $a_E$ decrease whereas $r^E_H$ increases with increase in $g$. The region between the event horizon and static limit surface is known as ergoregion from where energy can be extracted via the Penrose process \cite{Penrose:1971uk}. We show that as we increase the value of $g$ for corresponding $a$, then the region between the event horizon and static limit surface increases. In figures~\ref{fig1} and \ref{fig2}, we plot the contours of the horizon surfaces in the $xz$-plane of rotating regular Hayward's black hole to show the ergoregion for various values of $g$. The ergoregion for the rotating regular Hayward's black hole is shown in figure~\ref{fig2}, whereas the ergoregion of the Kerr black hole is shown in figure~\ref{fig1}. The ergoregion has two boundaries the event horizon and the outer static limit surface. The observer in ergoregion can't remain static. The ergoregion is sensitive to the deviation parameter $g$, and it's area increases with $g$. This can be seen by the event horizon $r_{H}^{+}$ and the static limit surface ($r^+_{sls}$) which are depicted in table~\ref{tab}, for various values of $a$ and $g$. Interestingly, $\delta^a=r^+_{sls}-r^+_H$, increases with $g$ as well as $a$, thereby suggesting that ergoregion enlarges. Further, we find that for each value of rotation parameter $a$, there exists a critical value of $g=g^*$, such that for $g<g^*$, we have two horizons suggesting nonextremal Hayward's black hole, whereas for $g=g^*$, one gets an extremal black hole where the two horizons coincide (cf. figure~\ref{fig2}), and for $g>g^*$, we obtain no horizons or no black hole.

\begin{table}
\begin{center}
\caption{\label{tab} Table for different values of $a$ and $g$ for rotating regular Hayward's black hole. $\delta$ is the region between static limit surface and event horizon($\delta^{a}=r^{+}_{sls}-r^{+}_{H}$).}
\resizebox{\textwidth}{!}{  
\begin{tabular}{| l | l l l | l l l | l l l | l l l |}
\hline
&\multicolumn{3}{c|}{$a=0.6$} &\multicolumn{3}{c|}{$a=0.7$}  &  \multicolumn{3}{c|}{$a=0.8$} &  \multicolumn{3}{c|}{$a=0.9$} \\
\hline
$g$  & $r^{+}_{H}$ & $r^{+}_{sls}$  & $\delta^{a}$ & $r^{+}_{H}$ & $r^{+}_{sls}$  & $\delta^{a}$ & $r^{+}_{H}$ & $r^{+}_{sls}$ &  $\delta^{a}$ & $r^{+}_{H}$ & $r^{+}_{sls}$ &  $\delta^{a}$\\
\hline
0  & 1.80000  & 1.90554  & 0.10554  & 1.71414  & 1.86891  & 0.15477  & 1.60000  & 1.82462  & 0.22462  & 1.43589  & 1.77136  &  0.33547  \\
0.1  & 1.79961  & 1.90523  & 0.10562  & 1.71367  & 1.86858  & 0.15491  & 1.59935  & 1.82426  & 0.22491  & 1.43477  & 1.77095  &  0.33618  \\
0.3  & 1.78944  & 1.89725  & 0.10781  & 1.70103  & 1.85992  & 0.15889  & 1.58186  & 1.81466  & 0.23280  & 1.40355  & 1.76004  &  0.35649  \\  
0.5  & 1.74839  & 1.86576  & 0.11737  & 1.64846  & 1.82553  & 0.17707  & 1.50329  & 1.77620  & 0.27291  &  -       & 1.71562  & -          \\
\hline   
\end{tabular}
}
\end{center}
\end{table}

\section{Equations of motion of the particle}
\label{eom}
Next, we calculate the equations of motion of a particle which will be necessary to determine the $E_{CM}$ of the two colliding particles for the rotating regular Hayward's black hole (\ref{metric}). Indeed, we are interested to study the radial motion of the particle falling from rest at infinity in the background of rotating regular Hayward's black hole.
Henceforth, we will restrict our discussion to the equatorial plane ($\theta = \pi /2$). The motion of the particle is determined by the Lagrangian 
\begin{equation}
\mathcal{L}=\frac{1}{2} g_{\mu \nu} \frac{d x^{\mu}}{d \tau}\frac{d x^{\nu}}{d \tau},
\end{equation}
where $\tau$ is an affine parameter along the geodesic. Also, we know that the metric (\ref{metric}) have two Killing vectors, i.e., $\xi^a=(\frac{\partial}{\partial t})^a$ and $\chi^a=(\frac{\partial}{\partial \phi})^a$, which are respectively, associated with two conserved quantities, energy $E$ and angular momentum $L$. Furthermore, we can write the geodesic equations for a particle in terms of these conserved quantities
\begin{eqnarray}\label{cqt}
g_{ab} \xi^a u^b &=& -E, \nonumber \\
g_{ab} \chi^a u^b &=& L, \nonumber
\end{eqnarray}
where $u^a=d x^a/d \tau$, represents a 4-velocity. The above equations can be written as
\begin{equation}\label{E}
g_{tt}u^{t}+g_{t \phi}u^{\phi}=-E,
\end{equation}
\begin{equation}\label{L}
g_{t \phi}u^{t}+g_{\phi \phi}u^{\phi}=L.
\end{equation} 
Solving eqs.~(\ref{E}) and (\ref{L}), simultaneously, and using $p_{\mu}p^{\mu}=-m^2 $, we obtain the geodesic equations
\begin{equation}\label{u^t}
 u^{t} = \frac{1}{r^2 \Delta} \Big[-a (aE - L)\Delta + \left(r^2 + a^2\right) \mathcal{P} \Big],
\end{equation}
\begin{equation}\label{u^Phi}
 u^{\phi} = \frac{1}{r^2 \Delta} \Big[-(aE - L ) \Delta+a \mathcal{P} \Big], 
\end{equation}
\begin{equation}\label{u^r}
u^{r} = \pm \frac{1}{r^2} \sqrt{\mathcal{P}^2 -\Delta \left[m^2 r^2  + (L-a E)^2  \right]},
\end{equation}
where $m$ corresponds to the mass of a particle, and $$\mathcal{P} = (r^2 + a^2)E -L a .$$ 
One can easily check that if we set $g=0$, then eqs.~(\ref{u^t}), (\ref{u^Phi}) and (\ref{u^r}) reduce to the  geodesic equations of the Kerr black hole. Thus the eqs.~(\ref{u^t}), (\ref{u^Phi}), and (\ref{u^r}) are corrected by the deviation parameter $g$, and in the limit $g \rightarrow 0$, we set corresponding equations for the Kerr black hole \cite{Banados:2009pr}. The radial equation for the timelike particles moving along the geodesics is described by
\begin{equation}
\frac{1}{2} (u^r)^2 + V_{eff} = 0,
\end{equation}
where $V_{ eff}$ is the effective potential given by
\begin{equation}
 V_{ eff} =  -\frac{[(r^2 + a^2)E -La]^2 -\Delta [m^2 r^2  + (L-a E)^2]}{2 r^4}.
\end{equation}
We need to study $V_{ eff}$ to get the range for the angular momentum with which the particle can approach the black hole. One can have an idea about the allowed and prohibited regions of the particle around the black hole, i.e., $V_{eff} \leq 0$ and $V_{eff}>0$. We can obtain the possible values of angular momentum of the test particle by using the circular orbits conditions, i.e.,
\begin{eqnarray}
\label{lim}
V_{ eff}=0 \;\;\;\text{and}\;\;\; \frac{dV_{ eff}}{dr}=0.
\end{eqnarray}
Here, we are interested to calculate the range of the critical angular momentum with which particle can reach the horizon of the black hole, which can be calculated from the effective potential using eq.~(\ref{lim}). Since, geodesics are timelike, i.e., $dt/d \tau \geq 0$, then eq.~(\ref{u^t}) leads to
\begin{eqnarray}
\frac{1}{r^2}[-a(aE-L)+(r^2+a^2)\frac{\mathcal{P}}{\Delta}]\geq 0,
\end{eqnarray}
the above condition at the horizon reduces to 
\begin{eqnarray}
E-\Omega_H L \geq 0, \nonumber\\ \Omega_H=\frac{a}{(r_H^E)^2+a^2}.
\end{eqnarray}
The critical angular momentum of the particle is defined by $L_C=E/\Omega_H$, where $\Omega_H$ is the angular velocity of the black hole at the horizon. This has been calculated numerically for the rotating regular Hayward's black hole in Ref. \cite{Amir:2015pja} for both extremal and nonextremal black holes. If $L > L_C$, then the particle will never reach to the horizon of the black hole, where $L_C$ is the critical angular momentum of the particle. Instead of it if $L < L_C$, then the particle will always be fall into the black hole, and if $L=L_C$, then the particle hit exactly at the event horizon of the black hole.

\section{Center-of-mass energy of the colliding particles in Hayward's black hole}
\label{cme}
We know the range of angular momentum, for which the particle can reach the horizon and collision takes place at the horizon of the black hole. Here, we shall study the $E_{CM}$ for the collision of two particles with rest masses $m_{1}$ and $ m_{2}$ ($m_{1}\neq m_{2}$), which  are initially at rest at infinity, moving towards the rotating regular Hayward's black hole and collide in the vicinity of the event horizon. Let us consider, the particles are coming with energy $E_1$, $E_2$ and angular momentum $L_{1}$, $L_{2}$. The 4-momentum of $i^{th}$ particle is defined as
\begin{equation}\label{p}
p^{\mu}_{i} = m_{i} u^{\mu}_{i},
\end{equation}
where $m_{i}$ and $ u^{\mu}_{i}$ correspond, respectively, to mass, and 4-velocity of the $i^{th}$ particle ($i=1,2$) and the total 4-momenta of the particles is given by
\begin{equation}\label{p_tot}
p^{\mu}_{t} = p^{\mu}_{(1)} + p^{\mu}_{(2)} = m_{1} u^{\mu}_{(1)} + m_{2} u^{\mu}_{(2)}.
\end{equation}
Then, $E_{CM}$ of the two particles is given by
\begin{equation}\label{ecfr}
E_{CM}^2 = - p^{\mu}_{t} p_{t \mu} = m_{1}^2 + m_{2}^2 - 2  g_{\mu \nu} p^{\mu}_{(1)} p^{\nu}_{(2)},
\end{equation}
by substituting eq.~(\ref{p_tot}) into the eq.~(\ref{ecfr}), we have
\begin{eqnarray}
E_{CM}^2 &=& m_{1}^2+m_{2}^2-2m_{1}m_{2}g_{\mu \nu}u^{\mu}_{(1)}u^{\nu}_{(2)},
\end{eqnarray}
which, due to $u^a u_a=-1$, can be rewritten in the following form
\begin{eqnarray}\label{ecfr1}
E_{CM} &=& \sqrt{2 m_1 m_2} \sqrt{1+\frac{(m_{1}-m_{2})^2}{2m_{1}m_{2}}-g_{\mu \nu}u^{\mu}_{(1)}u^{\nu}_{(2)}},
\end{eqnarray}
when $m_1=m_2=0$,
\begin{eqnarray}\label{ecfr2}
E_{CM} &=& m_0\sqrt{2} \sqrt{1-g_{\mu \nu}u^{\mu}_{(1)}u^{\nu}_{(2)}}.
\end{eqnarray}
The above formula (\ref{ecfr1}) is valid for both massless and massive particles. Inserting the values of $g_{\mu \nu}$,  $u^{\mu}_{(1)}$, and $u^{\nu}_{(2)}$ from eqs.~(\ref{metric}), (\ref{u^t}), (\ref{u^Phi}), (\ref{u^r}) into the eq.~(\ref{ecfr1}), we can obtain $E_{CM}$ of the two particles for the rotating regular Hayward's black hole
\begin{eqnarray}\label{masterecm}
\frac{E_{CM}^2}{2 m_1 m_2} &=& \frac{(m_{1}-m_{2})^{2}}{2 m_1 m_2} + \frac{1}{r(r^2-2 m(r) r+a^2)} \Big[a^2[(2 m(r)+r)E_1 E_2+r] \nonumber\\ 
&& -2a m(r)(L_{1} E_{2}+L_{2} E_{1}) -L_{1}L_{2} (-2m(r)+r)+[-2 m(r)+r(1+E_1 E_2)]r^2 \nonumber\\  
&& -\sqrt{r(r^2+a^2)(E^2_{1}-m_1^2) +2 m(r)(a E_{1}-L_{1})^2 - L_{1}^{2}r + 2 m(r) r^2 m_1^2 }
\nonumber \\  && 
\times \sqrt{r(r^2+a^2)(E^2_{2}-m_2^2) +2 m(r)(a E_{2}-L_{2})^2 - L_{2}^{2}r + 2 m(r) r^2 m_2^2}\Big],
\end{eqnarray}
where the mass $m(r)$ is given by eq.~(\ref{mas}). Obviously, the result eq.~(\ref{masterecm}) confirms that the parameter $g$ has an influence on $E_{CM}$. Thus, $E_{CM}$ depends on the parameter $g$ and $a$. When $g \rightarrow 0$, in eq.~(\ref{masterecm}), we obtain the $E_{CM}$ of two different mass particles of the Kerr black hole \cite{Harada:2011xz}
\begin{eqnarray}\label{ecm2}
\frac{E_{CM}^2}{2 m_{1} m_{2}}(g \rightarrow 0) &=& \frac{(m_{1}-m_{2})^{2}}{2 m_1 m_2} +\frac{1}{r(r^2-2 M r+a^2)} \Big[a^2[(2 M+r)E_1 E_2+r] \nonumber\\ 
&& -2a M(L_{1} E_{2}+L_{2} E_{1}) -L_{1}L_{2} (-2M + r)+[-2 M +r(1+E_1 E_2)]r^2
\nonumber\\ 
&& -\sqrt{r(r^2+a^2)(E^2_{1}-m_1^2)+2M(a E_{1}-L_{1})^2 - L_{1}^2 r + 2M r^2 m_1^2} \nonumber \\ 
&& \times \sqrt{r(r^2+a^2)(E^2_{2}-m_2^2)+2M(a E_{2}-L_{2})^2 - L_{2}^2r + 2 M r^2 m_2^2}\Big].
\end{eqnarray}
Furthermore, if we choose $m_{1}=m_{2}=m_{0}$, and $E_{1}=E_{2}=E=1$, then eq.~(\ref{masterecm}) reduces to 
\begin{eqnarray}\label{ec2}
\frac{E_{CM}^2}{2 m_0^2} &=&\frac{ 1}{r(r^2-2 m(r) r +a^2)}
\Big[2 a^2 (m(r) +r) -2am(r)(L_{1}+L_{2})- L_{1}L_{2} (-2m(r) +r) \nonumber\\ &&
+2(-m(r) +r)r^2 
 -\sqrt{2 m(r) (a-L_{1})^2 - L_{1}^{2} r +2 m(r) r^2} \nonumber \\ 
&& \times \sqrt{2 m(r) (a-L_{2})^2 - L_{2}^{2} r+2 m(r) r^2}\Big],
\end{eqnarray}
which represents the $E_{CM}$ of two equal mass particles as shown in \cite{Amir:2015pja}. Again, if we consider the case when $g \rightarrow 0$, $M=1$, $m_{1}=m_{2}=m_{0}$, and $E_{1}=E_{2}=E=1$, then eq.~(\ref{masterecm}) reduces to
\begin{eqnarray}\label{ec3}
\frac{E_{CM}^2}{2 m_0^2}(g \rightarrow 0) &=&\frac{1}{r(r^2-2 r +a^2)}\Big[2 a^2 (1+r)-2 a(L_{1}+L_{2})-L_{1}L_{2}(-2+r)+2(-1+r)r^2 \nonumber\\ 
&& - \sqrt{2 (a-L_{2})^2 -  L_{2}^{2} r + 2 r^2 }  \sqrt{2 (a-L_{1})^2 - L_{1}^{2} r + 2 r^2 } \Big],
\end{eqnarray} 
which is exactly the same expression as obtained in \cite{Banados:2009pr}. Thus, we have obtained $E_{CM}$ of the rotating regular Hayward's black hole for two different mass particles which is a generalization of the Kerr black hole. One can easily check that eq.~(\ref{masterecm}) has $0/0$-form at the event horizon of the black hole, therefore we apply the l'Hospital's rule to find out the limits of $E_{CM}$ at the event horizon of both extremal and nonextremal black holes one by one.  

\begin{figure}[tbp]
\centering
\includegraphics[width=.49\textwidth]{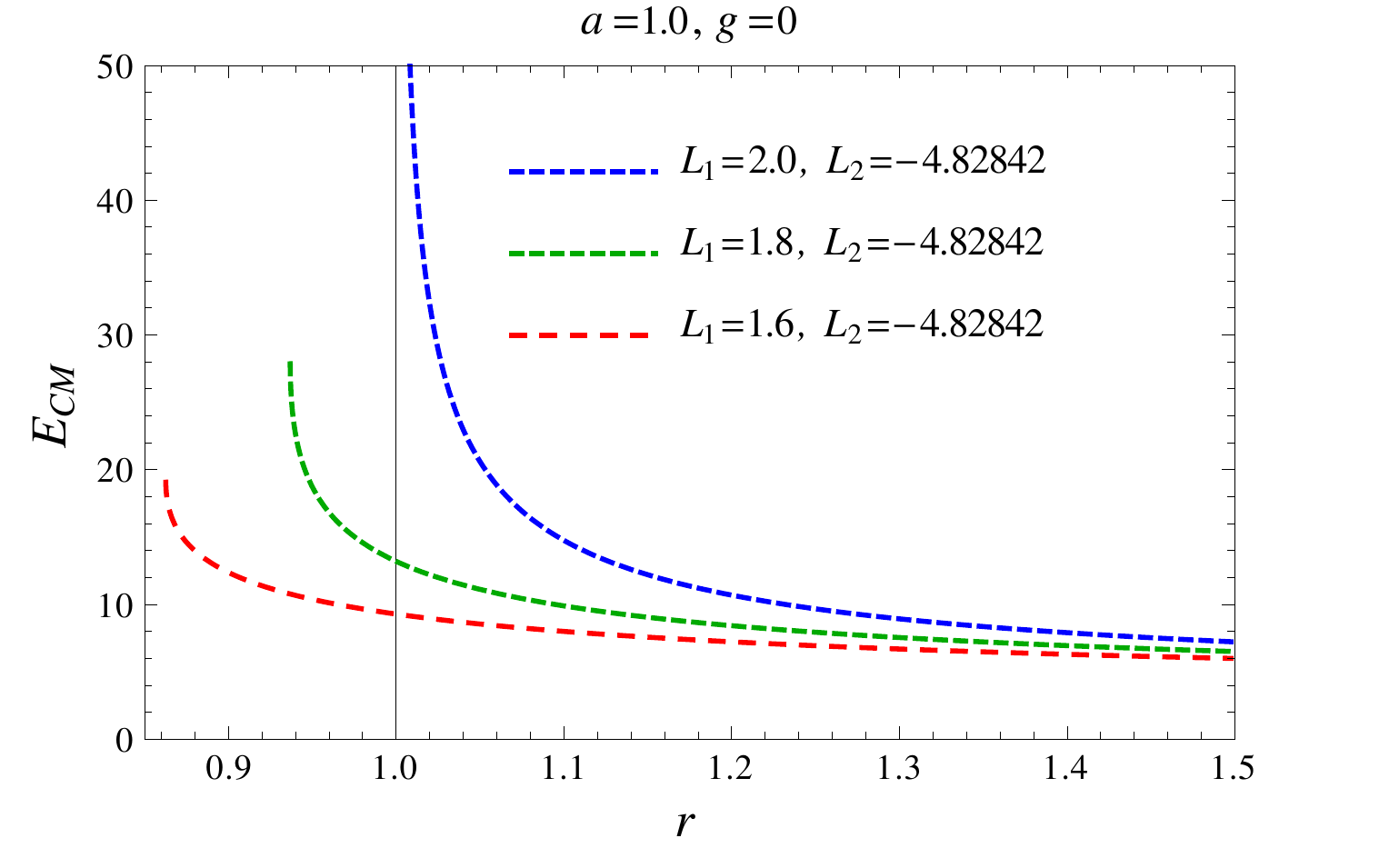}
\hfill
\includegraphics[width=.49\textwidth]{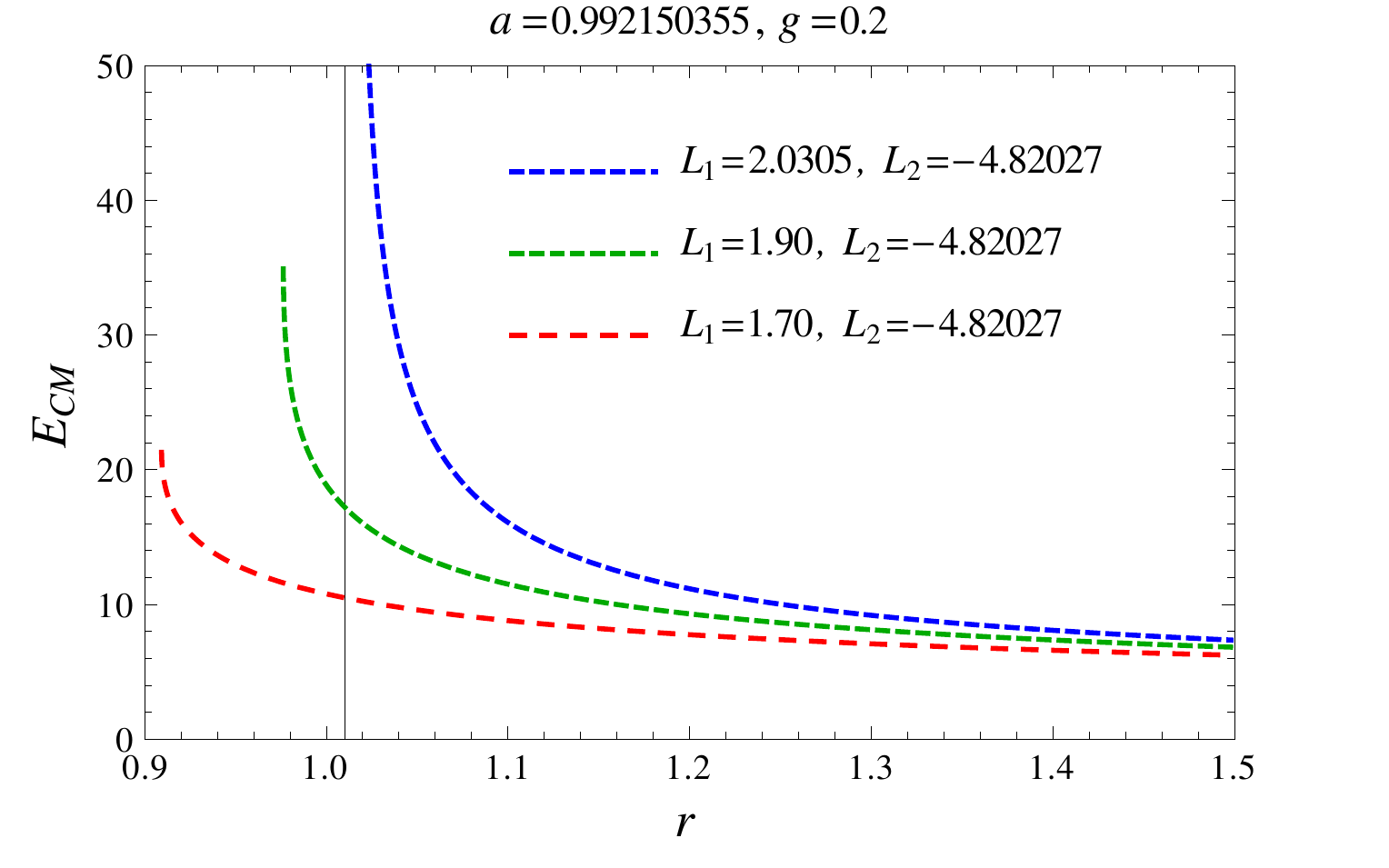}
\\
\includegraphics[width=.49\textwidth]{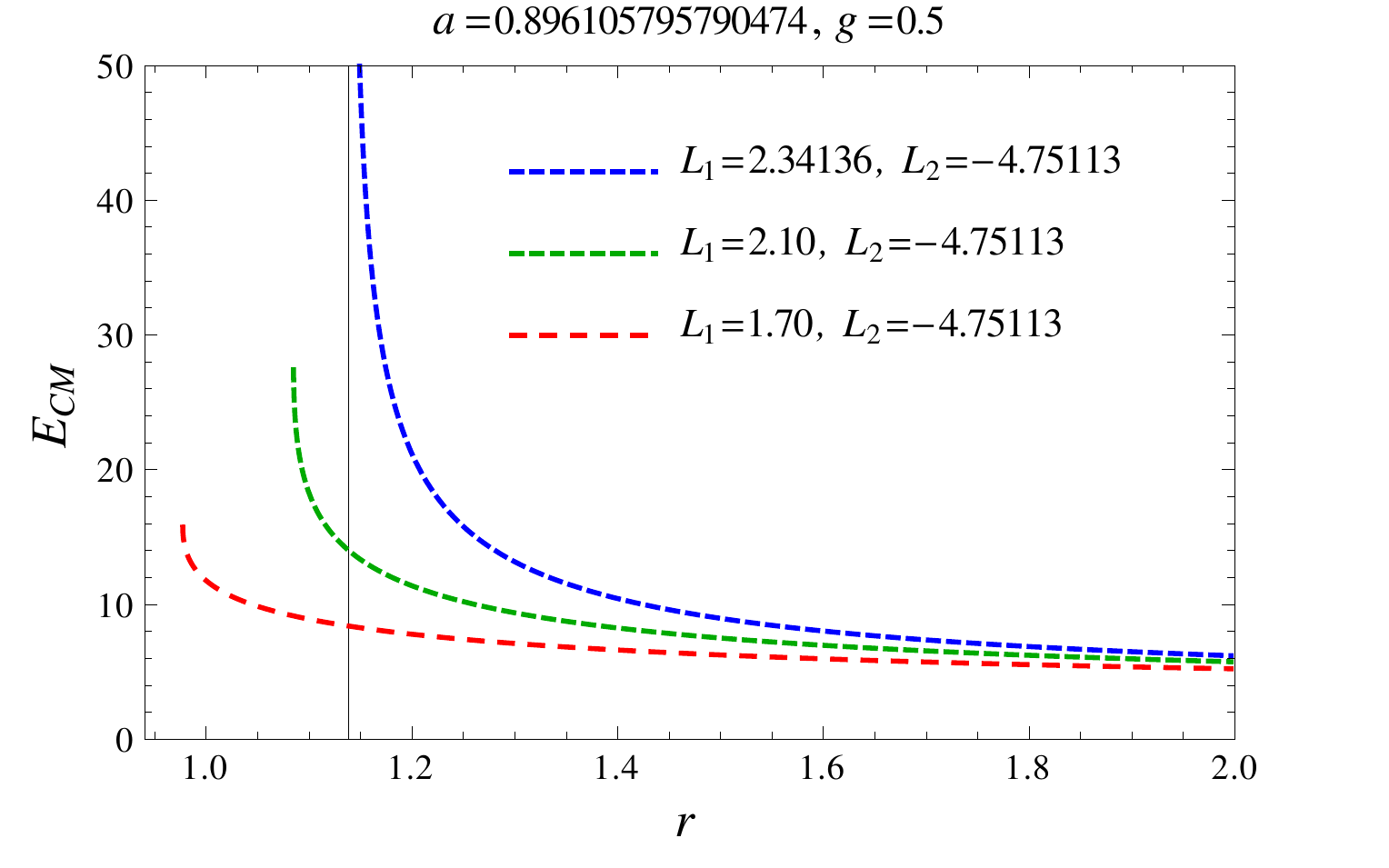}
\hfill
\includegraphics[width=.49\textwidth]{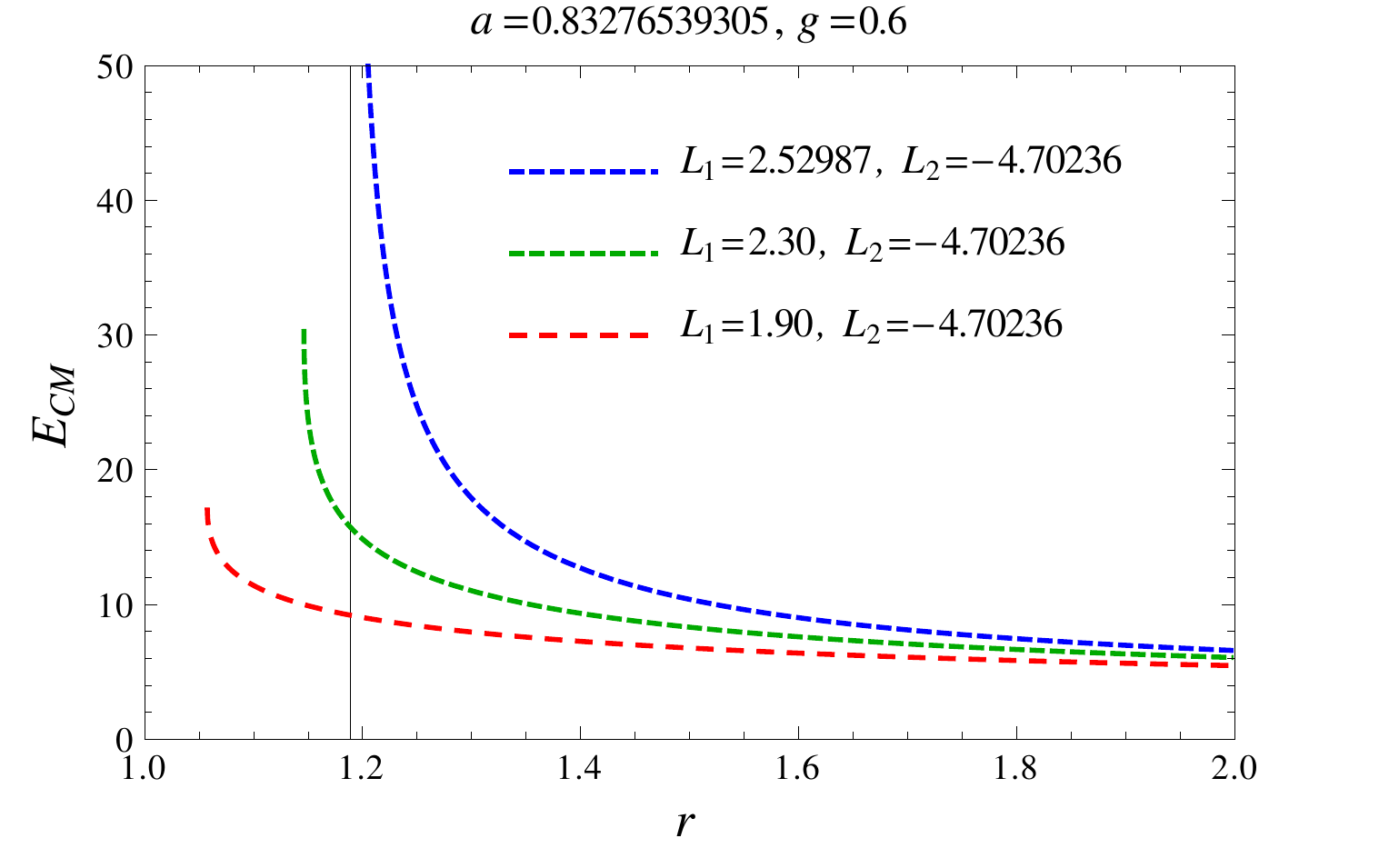}
\caption{\label{fig6} Plots showing the behavior of $E_{CM}$ vs $r$ for extremal rotating regular Hayward's black hole with mass of the colliding particles $m_{1}=1, m_{2}=2$ and different values of angular momentum $L_{1}$ and $L_{2}$. The vertical line corresponds to the location of event horizon.}
\end{figure}
\begin{figure}[tbp]
\centering
\includegraphics[width=.49\textwidth]{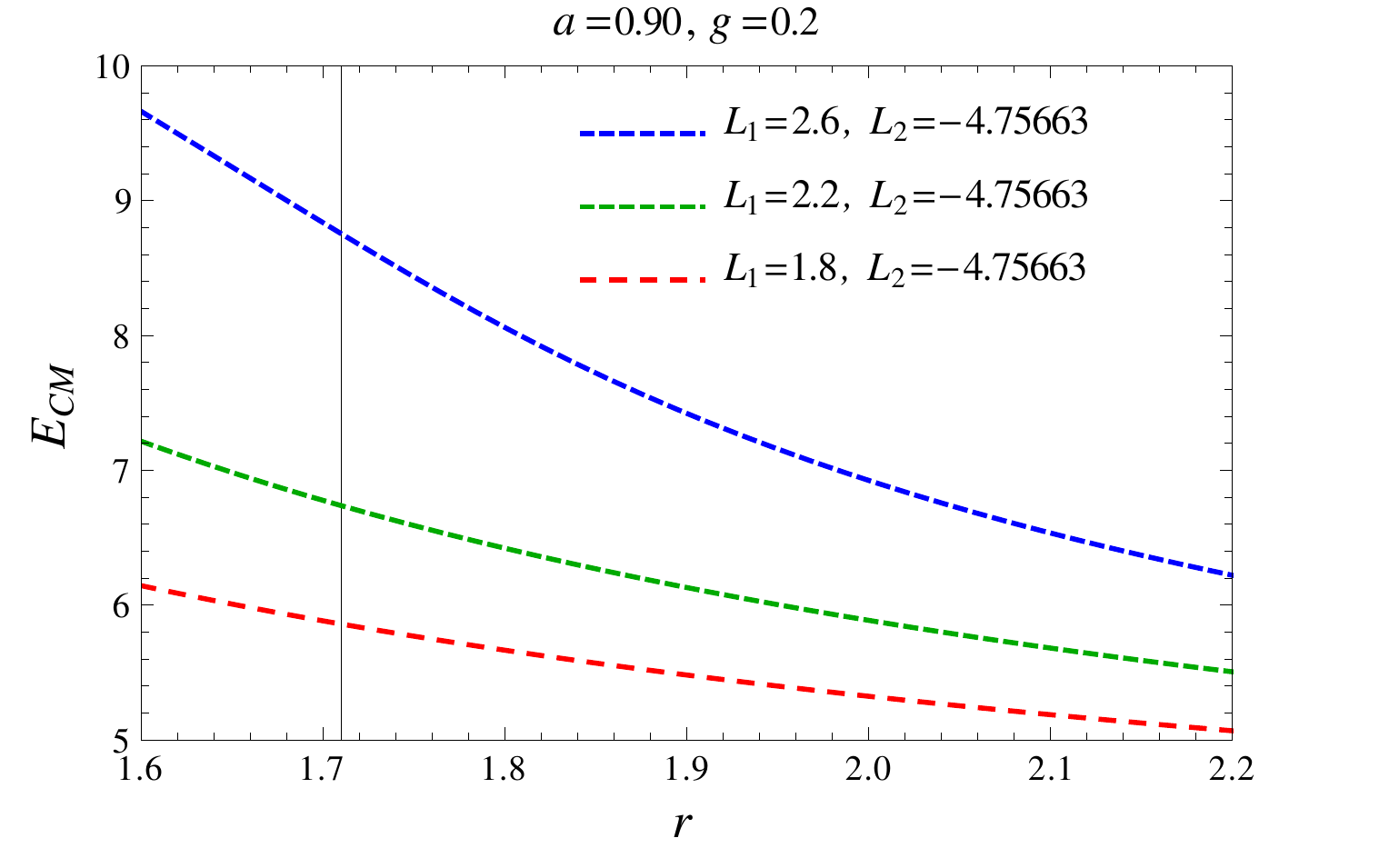}
\hfill
\includegraphics[width=.49\textwidth]{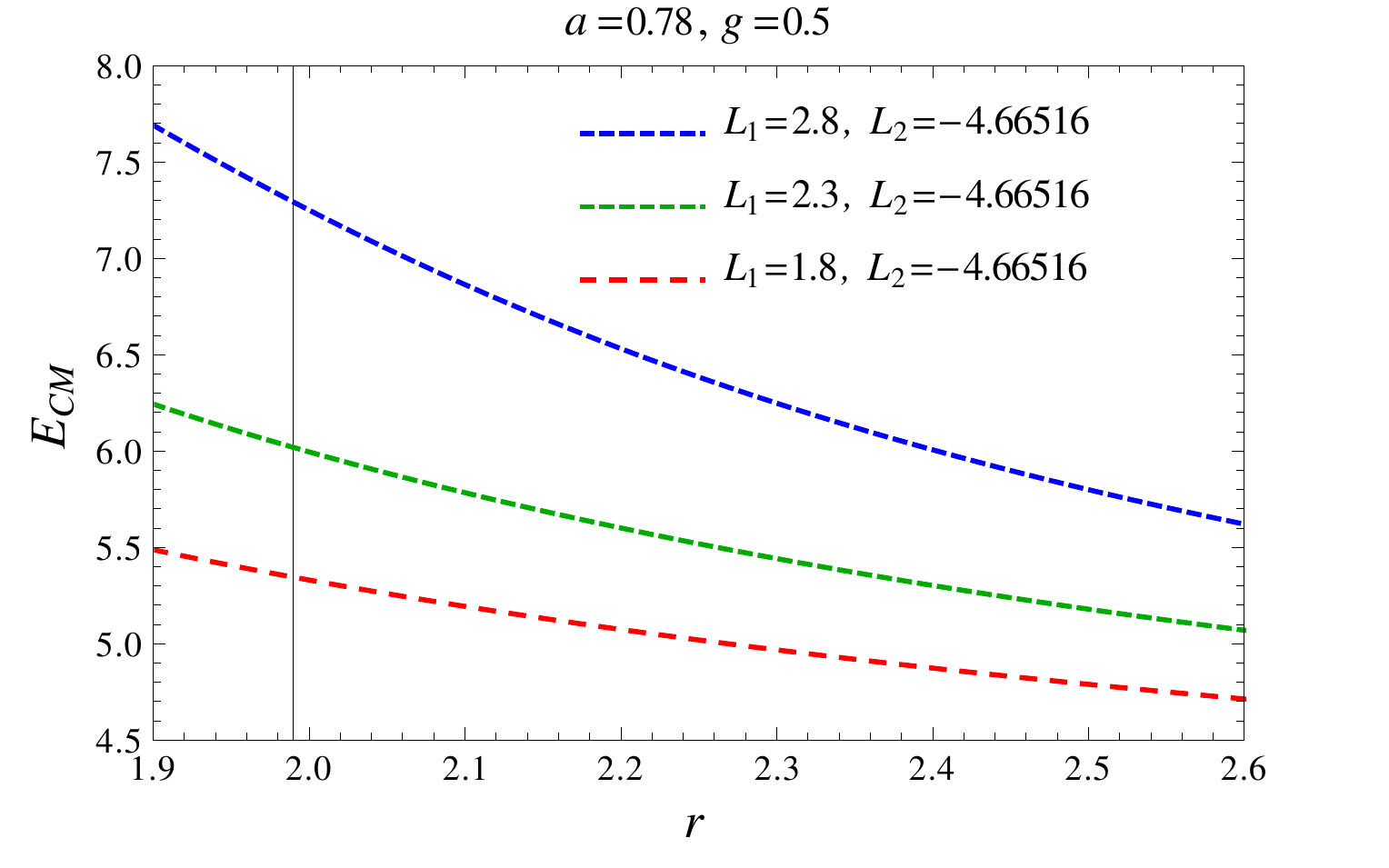}
\\
\includegraphics[width=.49\textwidth]{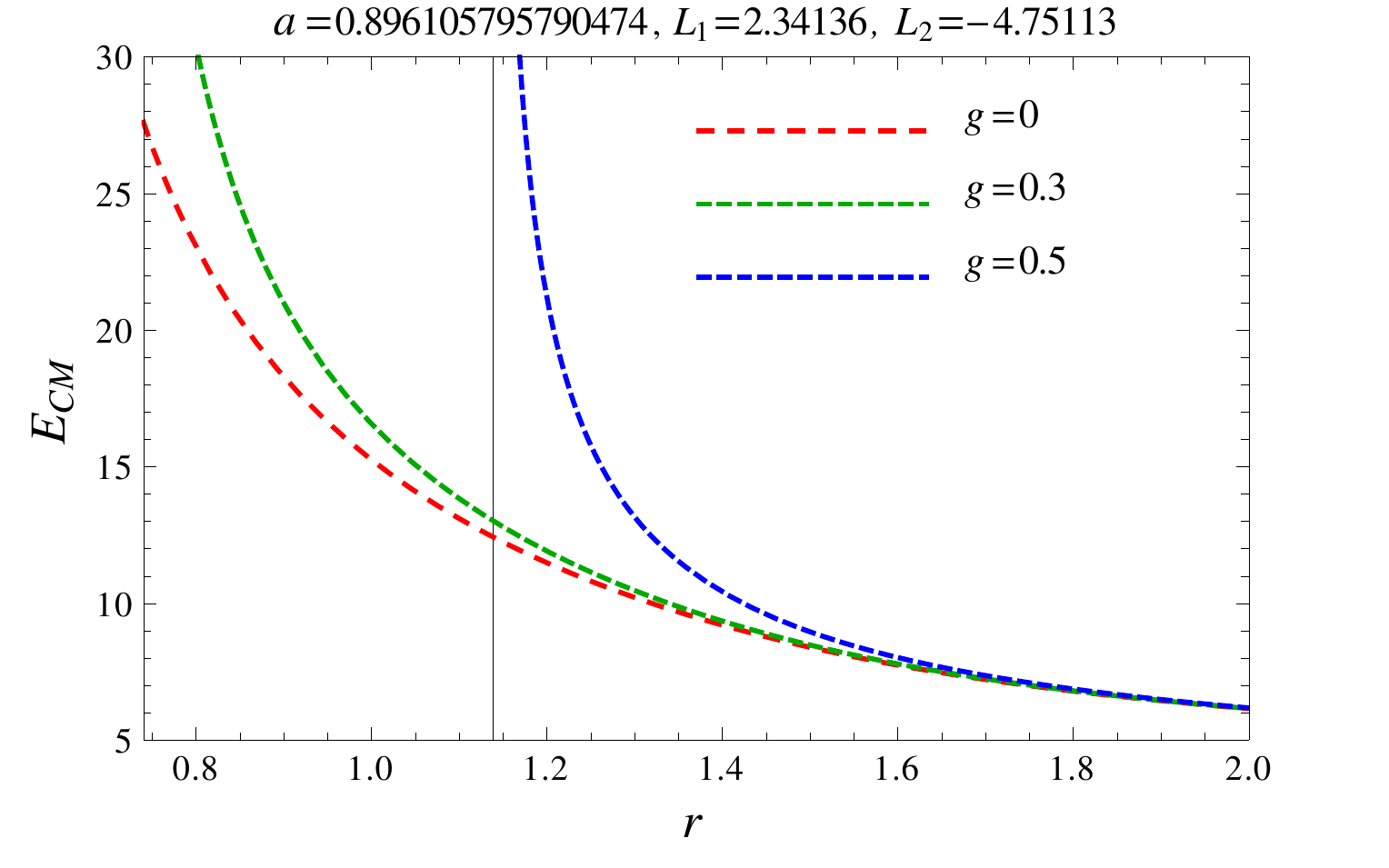}
\hfill
\includegraphics[width=.49\textwidth]{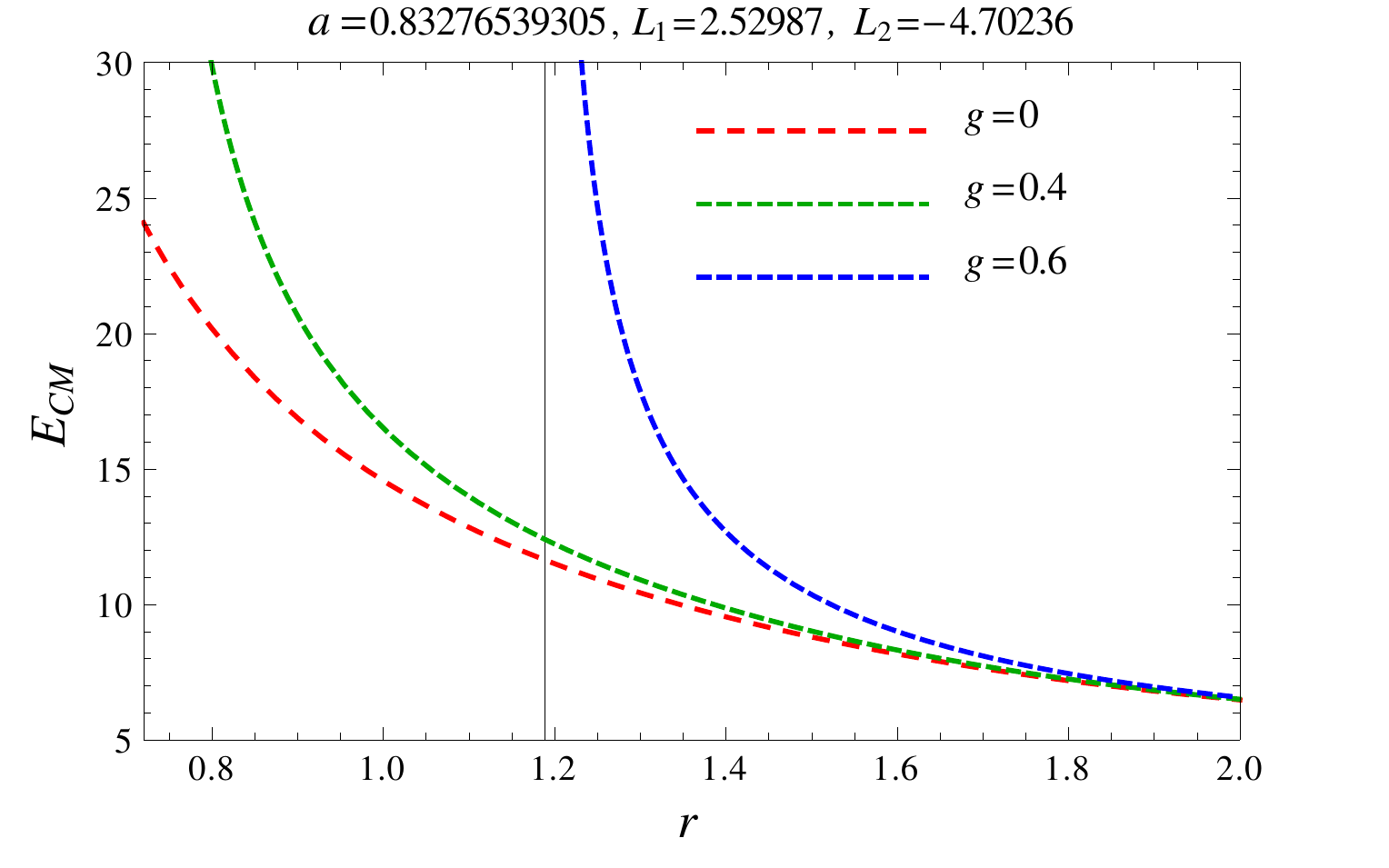}
\caption{\label{fig7} Plots showing the behavior of $E_{CM}$ vs $r$ for nonextremal rotating regular Hayward's black hole with mass of the colliding particles $m_{1}=1, m_{2}=2$, and for different values of $L_1$, $L_2$, and $g$. The vertical line corresponds to the location of event horizon.}
\end{figure}

\subsection{Near horizon collision in near-extremal Hayward's black hole}
Now, we can study the properties of $E_{CM}$~(\ref{masterecm}) as $r \rightarrow r_H^E$ of the extremal rotating regular Hayward's black hole. We note that in the limit $r \rightarrow r_H^E$, eq.~(\ref{masterecm}) has $0/0$-form. Hence, we apply the l'Hospital's rule twice and get the following form
\begin{eqnarray}\label{exta}
\frac{E_{CM}^2}{2m_{1}m_{2}}(r \rightarrow r^{E}_{H}) &=& \frac{(m_{1}-m_{2})^2}{2m_{1}m_{2}}
+\frac{1}{(E_1-\Omega_H L_1) (E_2-\Omega_H L_2)}\Big[0.60991 L_1 L_2\nonumber \\
&& + 1.67177\left[(E_1-\Omega_H L_1)^2 m_{2}^2+(E_2-\Omega_H L_2)^2 m_{1}^2\right] -1.42804(L_1 E_2+L_2 E_1)\nonumber \\
&& +0.49184(L_{1}^2 E_{2}^2+L_{2}^2 E_{1}^2)+(3.34356+0.98368 L_1 L_2)E_1 E_2\Big],
\end{eqnarray}
where, $a = a^E_H= 0.896105795790474$, $M=1$, $g=0.5$, $r^E_H=1.13802$. The critical values of the angular momentum can be calculated by $E_i-\Omega_{H}L_{i}=0$ or $L_{C_i}=E_i/\Omega_{H}$ $(i=1,2)$, where $L_{C_i}$ represents the critical value of angular momentum of $i^{th}$ particle and $\Omega_{H}=a/(r_H^2+a^2)$ is angular velocity at the event horizon of the black hole. It is clear from the eq.~(\ref{exta}) that if either $L_{C_1}=L_1$ or $L_{C_2}=L_2$, then the $E_{CM}$ becomes infinite, i.e., the necessary condition for getting $E_{CM}$ infinite is $L=L_C$ or $\Omega_H L =E$, for each particle. One can also see the behavior of $E_{CM}$ vs $r$ for an extremal rotating regular Hayward's black hole from figure~\ref{fig6}; it can be seen that $E_{CM}$ is infinite for the critical values of angular momentum $L_1=2.0, 2.0305, 2.34136, 2.52987$ corresponding to $g=0, 0.2, 0.5, 0.6$ and it remains finite for other values. 

Next we consider, the limits $E_ 1 = E_2 = E$, and $m_1=m_2=m_0$ in this case the eq.~(\ref{exta}) reduces to 
\begin{eqnarray}\label{ext1}
\frac{E_{CM}^2}{2m_0^2}(r \rightarrow r^{E}_{H}) &=& \frac{1}{(E-\Omega_H L_1) (E-\Omega_H L_2)}\Big[0.60991 L_1 L_2 \nonumber \\
&& + 1.67177\left[(E-\Omega_H L_1)^2 +(E-\Omega_H L_2)^2 \right]m_{0}^2 -1.42804(L_1+L_2) E \nonumber \\
&& +0.49184(L_{1}^2+L_{2}^2)E^2+(3.34356+0.98368 L_1 L_2)E^2\Big].
\end{eqnarray}
Here the critical angular momentum in this case can be calculated by using $L_{C}=E/\Omega_{H}$. One can observe from eq.~(\ref{ext1}) that if either $L_C=L_1$ or $L_C=L_2$, i.e., if one of the particle is coming with critical angular momentum, then we should get an infinite amount of $E_{CM}$ and the angular momentum of one of the particles is not equal to $L_C$ or greater than $L_C$, then the amount of $E_{CM}$ is finite. One can obtain the limit of $E_{CM}$ (\ref{masterecm}) for $g \rightarrow 0$ at $r \rightarrow r_H^E$, and $a=a_E$, which will take the following form
\begin{eqnarray}\label{ext2}
\frac{E_{CM}^2}{2m_{1}m_{2}}(r \rightarrow r^{+}_{H}) &=& \frac{(m_{1}-m_{2})^2}{2m_{1}m_{2}} 
+\frac{1}{4 (E_1-\Omega_H L_1) (E_2-\Omega_H L_2)} \Big[4 (E_1 -\Omega_H L_1)^2 m_2^2 \nonumber \\
&& +8 E_2 (E_1-\Omega_H L_1) -4 (E_1-\Omega_H L_1) L_2 +E_1^2 L_2^2 +E_2^2 L_1^2 -2 E_1 E_2 L_1 L_2  \nonumber \\
&& +4 E_2^2 m_1^2 -4 E_2 L_2 m_1^2 + L_2^2 m_1^2 \Big],
\end{eqnarray}
it represents the Kerr black hole case of two different massive colliding particles. 

\subsection{Particle collision for nonextremal black hole} 
If the two horizons of (\ref{metric}), do not coincide, we have a nonextremal rotating regular Hayward's black hole. As mention above that both numerator and denominator of eq.~(\ref{masterecm}) vanishes at $r \rightarrow r^{+}_H$. Hence, applying the l'Hospital's rule and calculate $E_{CM}$ at $r \rightarrow r^{+}_H$, after a tedious and like calculation, we obtain
\begin{eqnarray}\label{next}
\frac{E_{CM}^2}{2m_{1}m_{2}}(r \rightarrow r^{+}_{H}) &=& \frac{(m_{1}-m_{2})^2}{2m_{1}m_{2}} +\frac{0.5 (E_2-\Omega_H L_2)}{(E_1-\Omega_H L_1)} \Big[0.872182 L_1^2 +3.41504 E_1 L_1 -16.8792 E_1^2   \nonumber \\
&& + 1.97103 m_1^2\Big]+\frac{0.5 (E_1-\Omega_H L_1)}{(E_2-\Omega_H L_2)} 
\Big[0.872182 L_2^2 +3.41504 E_2 L_2-16.8792 E_2^2  \nonumber \\
&& +1.97103 m_2^2\Big] +2.13440 (0.8 E_1-L_1) (0.8 E_2-L_2) -3.00658 L_1 L_2 \nonumber \\
&&  + 1.97103 +15.51320 E_1 E_2,
\end{eqnarray}
where $a=0.8$, $g=0.5$ and $ r^{+}_{H}=1.50328$. The expression $E_i-\Omega_{H}L_{i}=0$ $(i=1,2)$ gives the critical angular momentum of a particle for nonextremal black hole case which is $L'_{C_i}=E_i/\Omega_{H}$. It seems from eq.~(\ref{next}) that either $L_{C_1}^{'}=L_1$ or $L_{C_2}^{'}=L_2$, then $E_{CM}$ diverges, but in this case it is not possible because $L_{C_1}^{'}$ and $L_{C_2}^{'}$ does not lie in the range of angular momentum. Hence, one can say that $E_{CM}$ never become infinite, it will remain finite. Furthermore, for $E_1 = E_2 =E$, and $m_1=m_2=m_0$ then eq.~(\ref{next}) takes the form
\begin{eqnarray}\label{next1}
\frac{E_{CM}^2}{2m_0^{2}}(r \rightarrow r^{+}_{H}) &=& \frac{0.5 (E-\Omega_H L_2)}{(E-\Omega_H L_1)} \left(0.872182 L_1^2 +3.41504 E L_1 -16.8792 E^2 + 1.97103 m_0^2\right)   \nonumber \\
&& +\frac{0.5 (E-\Omega_H L_1)}{(E-\Omega_H L_2)} 
\left(0.872182 L_2^2 +3.41504 E L_2-16.8792 E^2 + 1.97103 m_0^2\right)  \nonumber \\
&& +2.13440 (0.8 E-L_1) (0.8 E-L_2) -3.00658 L_1 L_2 + 1.97103 \nonumber \\
&& +15.51320 E^2, 
\end{eqnarray}
Figure \ref{fig7} shows the behavior of $E_{CM}$ with radius $r$ for a nonextremal black hole which indicates that $E_{CM}$ remains finite for each value of $L_{1}$ and $L_{2}$. Furthermore, the effect of $g$ can also be seen from the figure~\ref{fig7}, which indicates that the $E_{CM}$ increases with $g$. The limit of nonextremal $E_{CM}$ for two different mass particles, in the case of $g \rightarrow 0$ has the following form 
\begin{eqnarray}\label{next2}
\frac{E_{CM}^2}{2m_{0}^2}(r \rightarrow r^{+}_{H}) &=& \frac{(m_{1}-m_{2})^2}{2m_{1}m_{2}} -\frac{0.5 (E_2-\Omega_H L_2)}{E_1-\Omega_H L_1} \left(19.712 E_1^2-3.2 E_1 L_1-1.2 L_1^2-3.072 m_1^2\right)\nonumber \\
&& + 2 (0.8 E_1-L_1) (0.8 E_2-L_2)+ 0.6 L_2^2 - 3.2 L_1 L_2 + 1.6 E_2 L_2 + 18.432 E_1 E_2  \nonumber \\
&& -9.856 E_2^2 +1.536 m_2^2+3.072,
\end{eqnarray}
where $a=0.8$, and $ r^{+}_{H}=1.6$. From eq.~(\ref{next2}), it is clear that $E_{CM}$, become infinite if one of the angular momentum get critical value.

\section{Conclusion}
\label{conclusion}
The singularity theorems under fairly general conditions imply that  a sufficiently massive collapsing object will undergo continual gravitational collapse, resulting in the formation of a singularity. However, It is widely believed that  singularities do not exist in Nature, but that they are an artefact of general relativity. Hence, especially in the absence of well defined quantum gravity, the regular models without singularities received much attention. Also, the astrophysical black holes  may be different from the Kerr black holes predicted in general relativity \cite{Bambi:2011mj}, but the actual nature of these objects has still to be verified. The rotating regular Hayward's black hole can be seen one of the non-Kerr black hole metrics, which in Boyer\(-\)Lindquist coordinates is same as the Kerr black hole with $ M $ replaced by a mass function $ m(r) $, which has an additional parameter due to the magnetic charge $g$ and that Hayward's black hole reduces to the Kerr black hole in the absence of charge ($g=0$). Interestingly, for each nonzero values of $g$, the Hayward's black hole has extremal black with critical spin parameter $a^*$ with degenerate horizons \cite{Amir:2015pja}. In this paper, we have performed a detailed analysis to study ergoregion in the rotating regular Hayward's black hole \cite{Bambi:2013ufa}, and discuss the effect of the deviation parameter $g$ on the ergoregion. Our analysis reveals (cf. figure~\ref{fig2}), for each $a$ there exists a critical $g=g^*$ such that for $g >g^*$, the rotating regular Hayward's black hole has a disconnected horizon, and the horizons coincide when $g=g^*$, and two horizons when $g<g^*$. When ergoregion of rotating regular Hayward's black hole compared with the Kerr black hole (cf. figure~\ref{fig1}), we find that the ergoregion is sensitive to the deviation parameter $g$, and the ergoregion enlarges as the value of the deviation parameter increases.

We have also studied the collision of two particles with the different rest masses moving in the equatorial plane of the rotating regular Hayward's black hole, and calculate the $E_{CM}$ for these colliding particles. We obtain a general expression of $E_{CM}$ for the rotating regular Hayward's black hole, the $E_{CM}$ is calculated when the collision takes place near the horizon for both extremal and nonextremal black holes. It is demonstrated that $E_{CM}$ not only depends on the rotation parameter $a$ of the rotating regular Hayward's black hole, but also on the deviation parameter $g$. Further, for an extremal black hole, we show that arbitrary high $E_{CM}$ when the collision occurs near the horizon and one of the particle has critical angular momentum. We also calculate $E_{CM}$ for the nonextremal rotating regular Hayward's black hole, and found that $E_{CM}$ is finite with an upper bound which increases with an increase in parameter $g$.

\acknowledgments
M.A. acknowledges the University Grant Commission, India, for financial support through the Maulana Azad National Fellowship For Minority Students scheme (Grant No.~F1-17.1/2012-13/MANF-2012-13-MUS-RAJ-8679). S.G.G. would like to thank SERB-DST for Research Project Grant NO SB/S2/HEP-008/2014. We also thanks IUCAA for hospitality while a part of this work was done.

\end{document}